\documentclass[8.5pt,twoside,twocolumn]{article}
\usepackage[super,sort&compress,comma]{natbib} 
\usepackage[version=3]{mhchem}
\usepackage{sectsty}
\usepackage{balance} 
\usepackage{float}
\usepackage[usenames, dvipsnames]{color}
\usepackage{epstopdf}
\usepackage{lastpage}
\usepackage[format=plain,justification=raggedright,singlelinecheck=false,font=small,labelfont=bf,labelsep=space]{caption} 
\usepackage{fancyhdr}
\usepackage[switch,columnwise]{lineno}
\usepackage{array}
\usepackage{droidsans}
\usepackage{charter}
\usepackage[T1]{fontenc}
\usepackage[usenames,dvipsnames]{xcolor}
\usepackage{setspace}
\usepackage[compact]{titlesec}
\usepackage{hyperref}
\usepackage{textcomp}
\usepackage{units}
\usepackage{color}
\usepackage{upgreek}
\usepackage{graphicx}
\usepackage{times,mathptmx}
\usepackage{etoolbox}
\usepackage{amssymb}
\usepackage{amsmath}
\usepackage{bm}%

\begin{document}

\thispagestyle{plain}
\fancypagestyle{plain}{
\fancyhead[L]{\includegraphics[height=8pt]{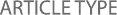}}
\fancyhead[C]{\hspace{-1cm}\includegraphics[height=20pt]{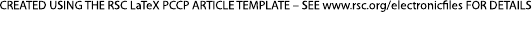}}
\fancyhead[R]{\includegraphics[height=10pt]{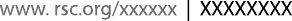}\vspace{-0.2cm}}
\renewcommand{\headrulewidth}{1pt}}
\renewcommand{\thefootnote}{\fnsymbol{footnote}}
\renewcommand\footnoterule{\vspace*{1pt}% 
\hrule width 3.4in height 0.4pt \vspace*{5pt}} 
\setcounter{secnumdepth}{5}

\makeatletter 
\def\subsubsection{\@startsection{subsubsection}{3}{10pt}{-1.25ex plus -1ex minus -.1ex}{0ex plus 0ex}{\normalsize\bf}} 
\def\paragraph{\@startsection{paragraph}{4}{10pt}{-1.25ex plus -1ex minus -.1ex}{0ex plus 0ex}{\normalsize\textit}} 
\renewcommand\@biblabel[1]{#1}            
\renewcommand\@makefntext[1]% 
{\noindent\makebox[0pt][r]{\@thefnmark\,}#1}
\makeatother 
\renewcommand{\figurename}{\small{Fig.}~}
\sectionfont{\large}
\subsectionfont{\normalsize} 

\fancyfoot{}
\fancyfoot[LO,RE]{\vspace{-7pt}\includegraphics[height=9pt]{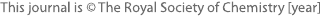}}
\fancyfoot[CO]{\vspace{-7.2pt}\hspace{12.2cm}\includegraphics{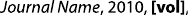}}
\fancyfoot[CE]{\vspace{-7.5pt}\hspace{-13.5cm}\includegraphics{headers/RF}}
\fancyfoot[RO]{\footnotesize{\sffamily{1--\pageref{LastPage} ~\textbar  \hspace{2pt}\thepage}}}
\fancyfoot[LE]{\footnotesize{\sffamily{\thepage~\textbar\hspace{3.45cm} 1--\pageref{LastPage}}}}
\fancyhead{}
\renewcommand{\headrulewidth}{1pt} 
\renewcommand{\footrulewidth}{1pt}
\setlength{\arrayrulewidth}{1pt}
\setlength{\columnsep}{6.5mm}
\setlength\bibsep{1pt}

\twocolumn[
  \begin{@twocolumnfalse}
\noindent\LARGE{\textbf{Colloidal Rod Dynamics under Large Amplitude Oscillatory Extensional Flow}} %$^\dag$
\vspace{0.6cm}

 \noindent\large{Steffen M. Recktenwald,\textit{$^{a}$}  Vincenzo Calabrese,\textit{$^{a,b}$} Amy Q. Shen,\textit{$^{a}$} Giovanniantonio Natale,\textit{$^{c}$} and Simon J. Haward$^{\ast}$\textit{$^{a}$}} \\\vspace{0.5cm}
%Please note that \ast indicates the corresponding author(s) but no footnote text is required. 

\noindent\textit{\small{\textbf{Received Xth XXXXXXXXXX 20XX, Accepted Xth XXXXXXXXX 20XX\newline
First published on the web Xth XXXXXXXXXX 200X}}}

\noindent \textbf{\small{DOI: 10.1039/b000000x}}
\vspace{0.6cm}
%Please do not change this text.

\noindent \normalsize{We perform a combined experimental and theoretical investigation of the orientational dynamics of rod-like colloidal particles in dilute suspension as they are subjected to a time-dependent homogeneous planar elongational flow. Our experimental approach involves the flow of dilute suspensions of cellulose nanocrystals (CNC) within a cross-slot-type stagnation point microfluidic device through which the extension rate is modulated sinusoidally over a wide range of amplitudes and periods, corresponding to a wide range of P\'{e}clet number amplitudes ($Pe_0$) and Deborah numbers ($De$), respectively. The time-dependent orientation of the CNC is experimentally assessed via quantitative flow-induced birefringence measurements using a high-speed polarization imaging camera. For small $Pe_0 \lesssim 1$ and small $De \lesssim 0.03$, the birefringence response is sinusoidal and in phase with the strain rate, \textit{i.e.}, the response is linear. With increasing $Pe_0$, the response becomes non-sinusoidal (\textit{i.e.}, nonlinear) as the birefringence saturates due to the high degree of particle alignment at higher strain rates during the cycle. With increasing $De$, the CNC rods have insufficient time to respond to the rapidly changing strain rate, leading to asymmetry in the birefringence response around the minima and a residual effect as the strain rate passes through zero. These varied dynamical responses of the rod-like CNC are captured in a detailed series of Lissajous plots of the birefringence versus the strain rate. Experimental measurements are compared with simulations performed on both monodisperse and polydisperse systems, with rotational diffusion coefficients $D_r$ matched to the CNC. A semiquantitative agreement is found for simulations of a polydisperse system with $D_r$ heavily weighted to the longest rods in the measured CNC distribution. The results will be valuable for understanding, predicting, and optimizing the orientation of rod-like colloids during transient processing flows such as fiber spinning and film casting.
}
\vspace{0.5cm}
 \end{@twocolumnfalse}
  ]

%Footnotes
\footnotetext{\textit{$^{a}$~Micro/Bio/Nanofluidics Unit, Okinawa Institute of Science and Technology Graduate University, 1919-1 Tancha, Onna-son, Okinawa 904-0495, Japan; E-mail: simon.haward@oist.jp}}

\footnotetext{\textit{$^{b}$~POLYMAT, Rheology and Advanced Manufacturing group, University of the Basque Country UPV/EHU, Avenida Tolosa 72, 20018, Donostia-San Sebastian, Spain}}

\footnotetext{\textit{$^{c}$~Department of Chemical and Petroleum Engineering, Schulich School of Engineering, University of Calgary, 2500 University Drive NW, Calgary, Alberta T2N 1N4, Canada}}

\section{Introduction}

Suspensions of anisotropic particles are prevalent in a wide range of applications, including industrial processing, biomedical engineering, drug delivery, and the development of advanced soft materials. The macroscopic properties of these systems, such as mechanical strength, optical anisotropy, electrical conductivity, and magnetic responsiveness, are often governed by the degree and nature of particle alignment.\cite{Leahy2017, Kumar2019} To tailor these properties for specific applications, it is essential to control the orientation of anisotropic particles during flow. Consequently, understanding the hydrodynamic alignment behavior of such particles is of central importance in the design and optimization of processing operations.

While industrial processes typically involve complex, mixed, and transient flows, much of the previous research has focused on the dynamics of colloidal rods under simpler conditions, such as steady shear or extensional flows.\cite{lang2019microstructural, Calabrese2022, Calabrese2023, Calabrese2024b,osawa2025regulating,rosen2018dynamic} It has been experimentally shown that the extent of rod alignment (in the absence of inertia) depends on the P\'eclet number \mbox{$Pe$}, a non-dimensional parameter that captures the ratio between the deformation rate \mbox{$|E|$} and the rotational Brownian diffusion coefficient \mbox{$D_r$}, \textit{i.e.}, \mbox{$Pe = |E|/D_r$}. Three general trends are observed as a function of \mbox{$Pe$}. For \mbox{$Pe \lesssim 1$}, the deformation rate is insufficient to overcome rotational Brownian diffusion, leading to rods that remain isotropically oriented, as in equilibrium conditions. At \mbox{$Pe \gtrsim 1$}, rods begin to spend an increasing fraction of their time in a preferentially oriented configuration, and as \mbox{$Pe \rightarrow \infty$}, they approach a limiting and maximal degree of alignment. \cite{Doi1978, Calabrese2023} Experimentally, the extent of rod alignment as a function of \mbox{$Pe$} has typically been probed using birefringence or small-angle scattering techniques, in combination with torsional rheometers to control shearing flow, or microfluidic devices to allow precise control over both flow type and strength. \cite{lang2019microstructural, ghanbari2025propagation, Calabrese2022, Calabrese2023, Calabrese2024b,detert2023alignment,Corona2018,lutz2016scanning,rodriguez2021nanostructure,corona2022fingerprinting}

In recent years, considerable interest has been devoted to understanding the dynamics of colloidal rods in complex flow environments that more closely resemble real-life conditions. For example, flow-focusing devices, contraction–expansion geometries, and fluidic four-roll mills have been employed to mimic specific flow fields of interest in a controlled manner.\cite{Corona2018,rodriguez2021nanostructure,lutz2016scanning,rosen2018dynamic,mittal2018multiscale,liu2020optimised} These devices share the common feature of generating regions with some degree of extensional flow, which is essential for replicating deformation modes relevant to many practical applications. In such systems, the flow is steady in time (\textit{i.e.}, in the \textit{Eulerian} frame), whereas particles experience a transient flow history, and consequently a varying deformation rate, as they are advected through the region of interest (\textit{i.e.}, in the \textit{Lagrangian} frame). Moreover, colloidal rods in these environments often undergo transient variations in flow type (\textit{e.g.}, during transitions from shear- to extension-dominated regions), which adds significant complexity, making data interpretation challenging and sometimes obscuring the key underlying physical mechanisms.

To overcome such difficulties while still capturing structural information relevant to real-life scenarios, one strategy is to study the dynamics of colloidal rods exposed to a constant type of deformation (or flow type) while transiently varying the magnitude of the deformation rate. In rheology, this is typically done with oscillatory shear tests. In particular, small-amplitude oscillatory shear (SAOS) has been established as a standard method for characterizing linear viscoelastic behavior under infinitesimal deformations, such that the material microstructure remains practically unperturbed from its equilibrium state.\cite{Ferry1947} However, with regard to measurements of structural evolution, SAOS provides little to no difference compared to equilibrium conditions, as a SAOS measurement, by definition, preserves the material’s equilibrium microstructure. In contrast, large-amplitude oscillatory shear (LAOS) is typically employed to probe the material response far from equilibrium, hence capturing features viscoelasticity and microstructural dynamics under nonlinear deformations.\cite{Hyun2002, Wilhelm2002,Khair_2016,Leahy2017,DeCorato2019,das2021shear} 

The presence of a rotational component in LAOS and shear experiments in general leads to dynamics that differ markedly from those observed in extensional flows. For instance, in shear flows, colloidal rods exhibit tumbling and wagging motions, types of dynamics that do not occur in purely extensional flows.\cite{vermant2001rheooptical,de2019oscillatory} Additionally, in shear, rods adopt a preferential orientation angle of \mbox{$\theta \approx 45^\circ$} relative to the flow direction at \mbox{$Pe \gtrsim 1$}, which progressively decreases toward \mbox{$\theta \rightarrow 0^\circ$} as \mbox{$Pe \rightarrow \infty$}. In contrast, in extensional flows, \mbox{$\theta \approx 0^\circ$} remains constant for \mbox{$Pe \gtrsim 1$}.\cite{reddy2018rheo,Calabrese2023,santos2023flow,Calabrese2021,vermant2001rheooptical,mohammad2022orientation} 

Perhaps even more striking is the case of flexible polymers, for which extensional flows induce a drastic conformational transition, from a coil-like state to a flow-sustained stretched configuration (the so-called coil-to-stretch transition), that is much less pronounced in shear flow.\cite{pope1978study, de1974coil,Calabrese2024b,smith1999single} This highlights the need to complement structural analyses performed under LAOS with its counterpart, large-amplitude oscillatory extension (LAOE), to investigate polymer and colloidal dynamics under transient extensional flows.\cite{Zhou2016, Zhou2016a, Recktenwald2025}

In our previous work, we demonstrated that LAOE under planar extension can be generated in an optimized microfluidic cross-slot, focusing on the retrieval of pressure drop and on the modification of the flow field induced by stretching polymers.\cite{Recktenwald2025} In this study, we extend the use of our LAOE setup to investigate the structural evolution of rigid colloidal rods, specifically cellulose nanocrystals (CNC), using birefringence techniques to probe the extent of rod alignment.
Rigid rods, such as the CNC employed here, not only serve as suitable building blocks for engineering well-defined, ordered soft bio-materials but also provide a useful model system for studying inelastic fluids governed solely by particle orientation. This stands in contrast to flexible polymers, which undergo complex flow-induced conformational changes, such as coil-to-stretch and stretch-to-coil transitions, under extensional flows. Nevertheless, the dynamic behavior of rod-like colloidal systems in time-dependent or oscillatory extensional flows remains largely unexplored.\cite{Kumar2023}

In this work, we employ the LAOE framework to study the dynamics of CNC as a model system for rigid colloidal rods. We demonstrate the usefulness of LAOE for investigating colloidal rod dynamics in transient extensional flows and provide a direct comparison with simulations to validate and support our experimental methodology. We reveal how the dynamical response of the CNC under LAOE depends on the amplitude and period of the applied sinusoidal extension rate, becoming strongly nonlinear at high amplitudes and short periods. We anticipate the results to be valuable for the optimization of particle alignment in processing applications involving transient mixed shear and extensional flows.

\section{Material and Methods}
\subsection{Test fluids and rheological characterization}
Cellulose nanocrystals (CNC, CelluForce, Montreal, Canada) are dispersed at volume concentrations (vol/vol) of \mbox{$\unit[0.05]{\%}$}, \mbox{$\unit[0.1]{\%}$}, and \mbox{$\unit[0.2]{\%}$} in a glycerol/water mixture containing \mbox{$\unit[80]{vol.\%}$} glycerol. This solvent mixture is a Newtonian fluid with a shear viscosity of \mbox{$\eta_{\mathrm{s}}=\unit[74]{mPa\,s}$}. The used CNC concentrations are chosen to be below the onset of interparticle interactions to ensure a marginal variation of the flow properties from those of a Newtonian fluid and an isotropic distribution of the rods at equilibrium.\cite{Bertsch2017, Bertsch2019, Calabrese2021} The rotational diffusion coefficient for non-interacting rods can be described as:\cite{Doi1978, Doi1988}

\begin{equation}
    D_r=\frac{3k_{\mathrm{B}}T\ln\left(l_{\mathrm{c}}/d_{\mathrm{eff}}\right)}{\pi\eta_{\mathrm{s}} {l_{\mathrm{c}}}^3},
    \label{eq_Dr} 
\end{equation}

\begin{figure}[ht]
\centering
\includegraphics[width=0.45\textwidth]{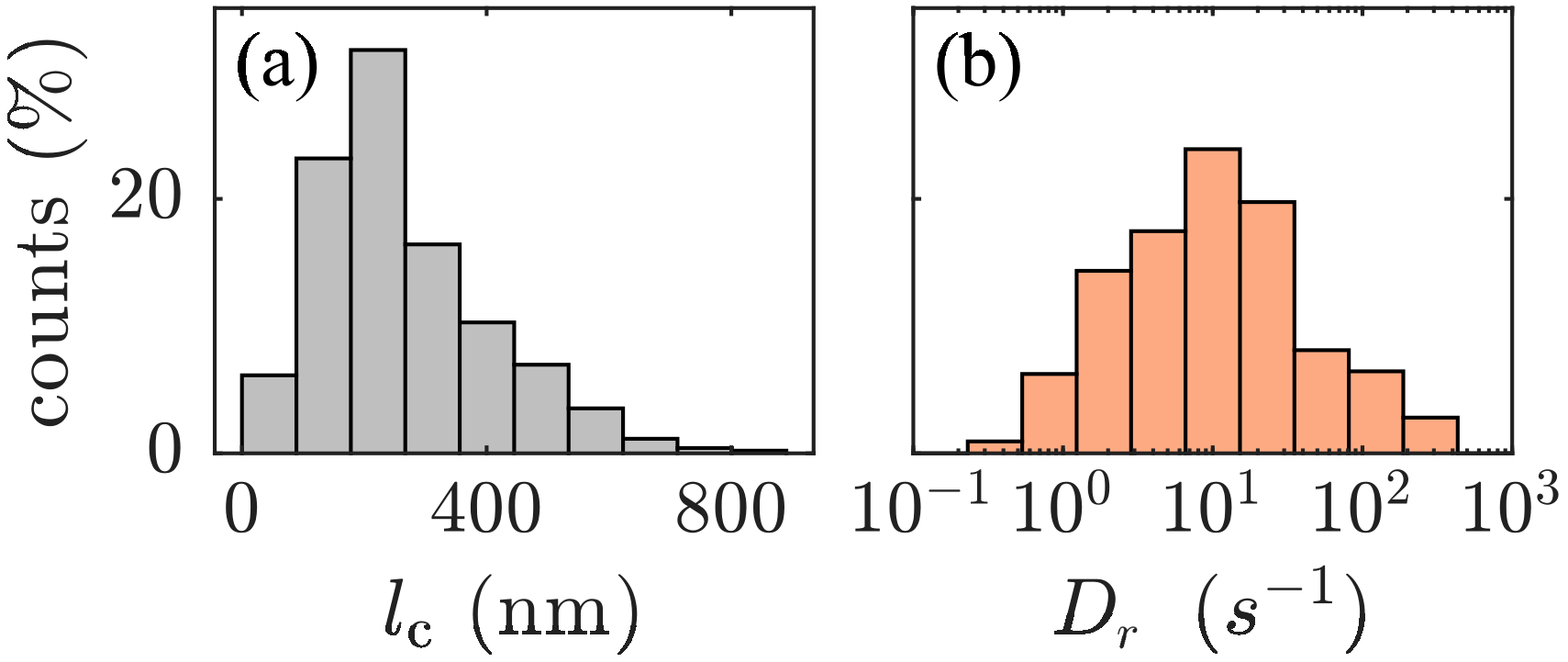}
\caption{CNC characterization. (a) Histogram of the characteristic contour length \mbox{$l_{\mathrm{c}}$} distribution using \mbox{$N=10$} bins, as obtained from AFM counting of 976 isolated particles (see Ref.~\citenum{Calabrese2021}). (b) Distribution of the rotational diffusion coefficient \mbox{$D_r$}, calculated from the data in (a) using Eq.~\ref{eq_Dr}, with $l_{\mathrm{c}}$ corresponding to that of each bin.}
\label{FIG_Dist} 
\end{figure}

\noindent where \mbox{$k_{\mathrm{B}}$} is the Boltzmann constant, \mbox{$T=\unit[ 295]{K}$} is the absolute temperature, \mbox{$\eta_{\mathrm{s}}$} is the solvent shear viscosity, \mbox{$l_{\mathrm{c}}$} is a characteristic contour length, and \mbox{$d_{\mathrm{eff}}$} is the effective diameter of the rods, which accounts for the bare rod diameter $d$ and the contribution of the electric double layer \mbox{$\delta d$} as $d_{\mathrm{eff}}=d+ \delta d$.\cite{Bertsch2019} We consider \mbox{$\delta d = \unit[22.6]{nm}$}, as reported in deionized water, to be a representative value of the electric double layer thickness.\cite{Bertsch2019} The CNC system used in this study has been thoroughly characterized previously.\cite{Calabrese2021, Calabrese2024b} Figure~\ref{FIG_Dist}(a) shows the size distribution obtained from atomic force microscopy (AFM), based on measurements made by Calabrese \textit{et al.}\cite{Calabrese2021} The rods exhibit a number-averaged contour length of \mbox{$\langle l_{\mathrm{c}}\rangle=\unit[260]{nm}$}, an average diameter of \mbox{$\langle d\rangle=\unit[4.8]{nm}$}, and a persistence length of \mbox{$l_{\mathrm{p}}=30\langle l_{\mathrm{c}}\rangle$}.\cite{Calabrese2021} Using \mbox{$\langle l_{\mathrm{c}}\rangle$}, an effective diameter \mbox{$d_{\mathrm{eff}}=\langle d\rangle+\delta d = \unit[27.4]{nm}$}, and the Newtonian solvent viscosity, the rotational diffusion coefficient is \mbox{$D_r \approx \unit[6.73]{s^{-1}}$} at \mbox{$\unit[22]{^\circ C}$}. Based on the volume-averaged length (\mbox{$\langle \overline{l_{\mathrm{c}}} \rangle=\unit[470]{nm}$}), we obtain \mbox{$\overline {D}_r \approx \unit[1.44]{s^{-1}}$}. The distribution of \mbox{$D_r$} (Fig.~\ref{FIG_Dist}(b)) used in the polydisperse simulations to be described later (Sec.~\ref{sec_methodsim}) is derived from the contour length distribution shown in Fig.~\ref{FIG_Dist}(a).

\begin{figure}[ht!]
\centering
\includegraphics[width=0.45\textwidth]{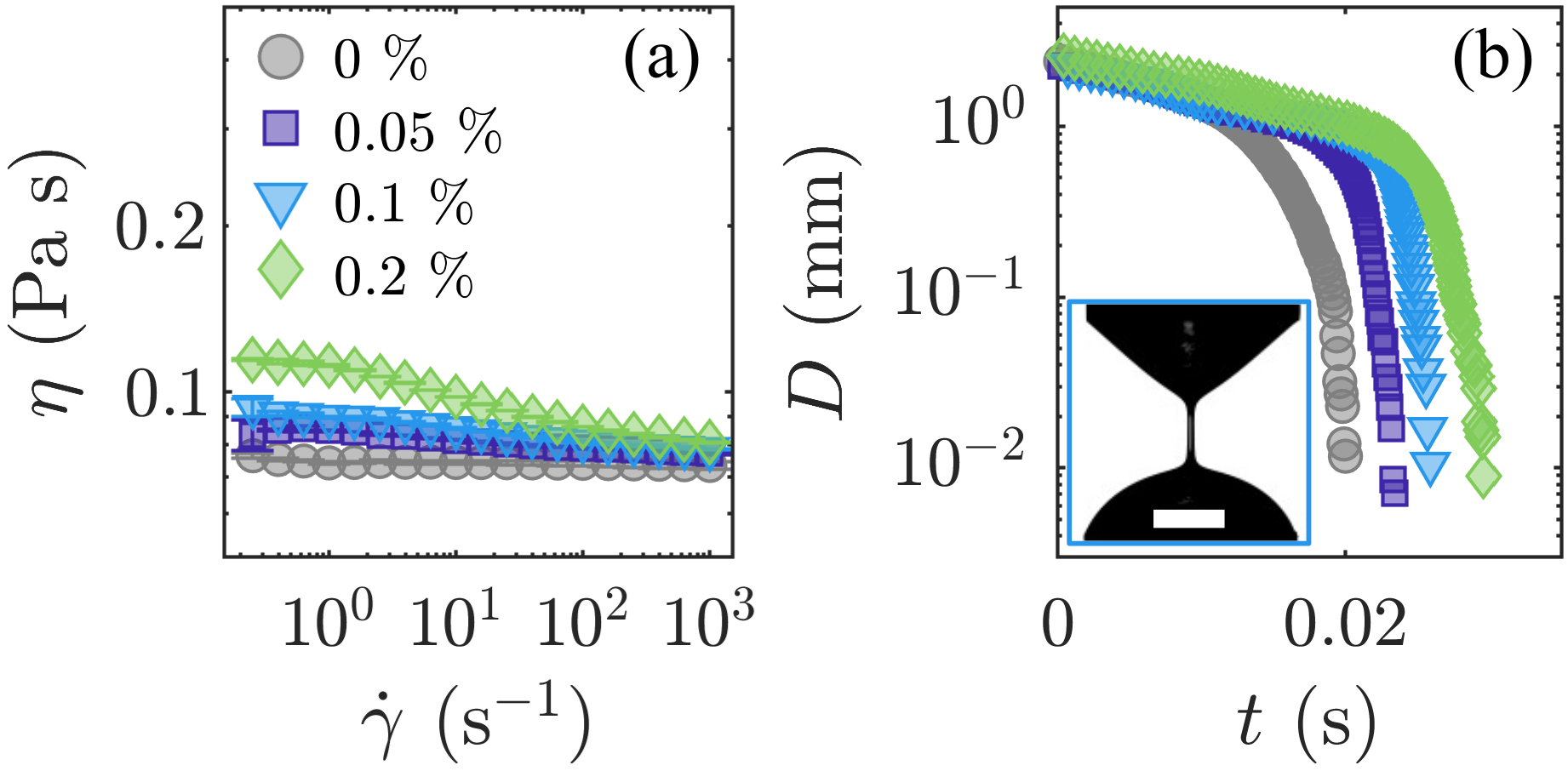}
\caption{Rheological responses of the CNC test fluids in shear and uniaxial extension. (a) Shear viscosity as a function of the applied shear rate. (b) Representative curves of the thinning filament diameter during capillary thinning in a CaBER experiment. The legend in (a) indicates the CNC concentration in vol.\%. The inset in (b) displays a representative snapshot of the filament for the \mbox{$\unit[0.1]{\%}$} CNC suspension, where the white scale bar represents \mbox{\unit[2]{mm}}.}
\label{FIG_Rheology} 
\end{figure}

The steady shear rheology of the test fluids was measured at \mbox{$\unit[22]{^\circ C}$} using a stress-controlled rotational rheometer (MCR502, Anton Paar, Austria) equipped with a double gap Couette geometry. %(DG27, Anton Paar, Austria). 
The \mbox{$\unit[0.05]{\%}$} and \mbox{$\unit[0.1]{\%}$} samples exhibit a nearly constant viscosity close to that of the Newtonian solvent (labeled as \mbox{$\unit[0]{\%}$}, see Fig.~\ref{FIG_Rheology}(a)). At \mbox{$\unit[0.2]{\%}$}, the sample shows slight shear-thinning behavior within the investigated shear rate range. Average viscosity values for the \mbox{$\unit[0.05]{\%}$}, \mbox{$\unit[0.1]{\%}$}, and \mbox{$\unit[0.2]{\%}$} CNC suspensions in the shear rate range of \mbox{$10 \leq \dot{\gamma} \leq \unit[1000]{s^{-1}}$} are \mbox{$\eta = \unit[79]{mPa\,s}$}, \mbox{$\eta = \unit[82]{mPa\,s}$}, and \mbox{$\eta = \unit[88]{mPa\,s}$}, respectively.

The behavior of the test fluids under uniaxial extension was characterized using a capillary breakup extensional rheometer (CaBER) device (Thermo-Haake, Germany). Circular plates with a diameter of \mbox{$\unit[6]{mm}$} were separated from an initial gap of \mbox{$\unit[2]{mm}$} to a final gap of \mbox{$\unit[6]{mm}$} within a strike time of \mbox{$\unit[20]{ms}$}. The diameter $D(t)$ of the resulting fluid filament was monitored as it thinned over time (see Fig.~\ref{FIG_Rheology}(b), where $t=0~\text{s}$ corresponds to the time at which the plates reached their final separation). We did not observe a detectable exponential filament decay characteristic of the elasto-capillary thinning regime, which is consistent with the rigid CNC rods being effectively inextensible and having negligible entropic elasticity.\cite{Anna2001a,Calabrese2024}

\subsection{Microfluidic setup}
\begin{figure}[ht]
\centering
\includegraphics[width=0.45\textwidth]{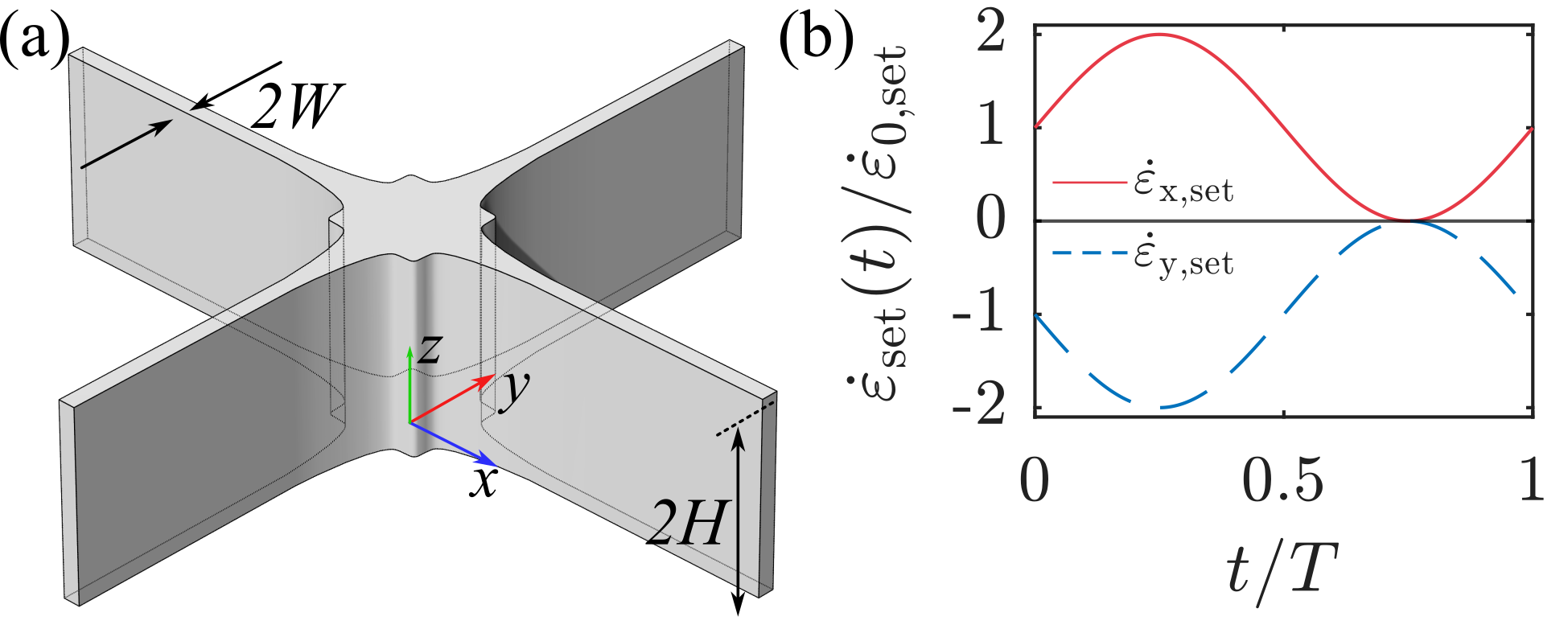}
\caption{Overview of the experimental setup. (a) Schematic illustration of the optimized shape cross-slot extensional rheometer (OSCER) geometry. The geometry has a height of \mbox{$2H$} and features two pairs of opposing inlet and outlet channels aligned along the \mbox{${x}$} and \mbox{${y}$} axes, respectively, each with a width of \mbox{$2W$}. (b) Representation of the normalized set extension rate profiles \mbox{$\dot{\varepsilon}_{\mathrm{x,set}}/\dot{\varepsilon}_{\mathrm{0,set}}$} and \mbox{$\dot{\varepsilon}_{\mathrm{y,set}}/\dot{\varepsilon}_{\mathrm{0,set}}$} in the \mbox{${x}$} and \mbox{${y}$} directions, respectively, shown over one normalized period \mbox{$t/T$}.}
\label{FIG_Setup} 
\end{figure}

A microfluidic optimized shape cross-slot extensional rheometer (OSCER) geometry was used to generate a planar extensional flow.\cite{Alves2008a, Haward2013a} The OSCER has a half-height of \mbox{$\unit[H=1]{mm}$} and a characteristic half-width of \mbox{$\unit[W=100]{\upmu m}$} for the inlet and outlet channels (Fig.~\ref{FIG_Setup}(a)). The high aspect ratio of the geometry \mbox{$H/W=10$} ensures a uniform flow across most of the channel's height and hence a good approximation to a two-dimensional (2D) planar flow field. The flow in the central part of the OSCER geometry (\mbox{$\lvert x/W \rvert,\lvert y/W \rvert \leq 15$}) approximates pure planar homogeneous elongation with hyperbolic streamlines surrounding a free stagnation point located at \mbox{$x=y=0$}.\cite{Haward2012d, GalindoRosales2014, Haward2016a} The geometry is fabricated from stainless steel using wire electrical discharge machining and is sealed with glass slides on both the top and bottom surfaces, enabling optical interrogation of the flow in the region surrounding the stagnation point.

\subsection{Flow control}
\subsubsection{Steady flow conditions}
The fluids are driven through the microfluidic channel using low-pressure syringe pumps (Nemesys S, \mbox{$29:1$} gear ratio, Cetoni GmbH, Germany), equipped with proprietary \mbox{$\unit[5]{mL}$} borosilicate glass syringes. The syringes are connected to the OSCER device via PTFE tubing (inner diameter \mbox{$\unit[1]{mm}$}, Darwin Microfluidics, France). Careful measures are implemented to eliminate air bubbles in the microfluidic system, which can impact the response of the system to the transient flows imposed during LAOE.\cite{Recktenwald2025}

Two of the pumps are set to inject the fluid at equal volumetric flow rates \mbox{$Q$} into the opposing inlet channels aligned along the \mbox{${y}$} direction, and two additional pumps are set to simultaneously withdraw fluid at equal and opposite rates from the outlet channels along the \mbox{${x}$} direction. The average flow velocity in each channel of the geometry is \mbox{$U = Q/(4WH)$} and the resulting set extension rate is \mbox{$\dot{\varepsilon}_{\mathrm{set}}=0.1U/W$}.\cite{Haward2012d, Haward2013a, Haward2016a,Haward2023b} Under this arrangement, the \mbox{${x}$} axis corresponds to the extensional axis and the \mbox{${y}$} axis is the compressional axis, and we denote the average strain rates along the extensional and compressional axes as \mbox{$\dot{\varepsilon}_{\mathrm{x}}$} and \mbox{$\dot{\varepsilon}_{\mathrm{y}}$}, respectively. For a Newtonian fluid under steady 2D planar elongation, \mbox{$\dot{\varepsilon}_{\mathrm{x}} = -\dot{\varepsilon}_{\mathrm{y}} = \dot{\varepsilon}_{\mathrm{set}}$}. However, for transient LAOE flows in the OSCER device, system damping can result in a phase shift and reduced amplitude of \mbox{$\dot{\varepsilon}_{\mathrm{x}}$} and \mbox{$\dot{\varepsilon}_{\mathrm{y}}$} relative to  \mbox{$\dot{\varepsilon}_{\mathrm{set}}$}.\cite{Recktenwald2025}

\subsubsection{Time-dependent flow conditions}
In this study, we investigate the response of our dilute CNC suspensions to pulsatile LAOE.\cite{Recktenwald2025} For this, we impose sinusoidal extension rate profiles at the inlets and outlets of the microfluidic OSCER device described by:

\begin{equation}
    \dot{\varepsilon}_{\mathrm{set}}(t) =\dot{\varepsilon}_{\mathrm{x,set}}(t) = -\dot{\varepsilon}_{\mathrm{y,set}}(t) = \dot{\varepsilon}_{\mathrm{{0,set}}} \left[1 + \sin\left(2\pi\, t/T\right)\right], 
    \label{eq_setsin} 
\end{equation}

\noindent where \mbox{$\dot{\varepsilon}_{\mathrm{{0,set}}}$} is the amplitude of the extension rate oscillation, and \mbox{$T=1/f$} is the pulsation period with \mbox{$f$} as the pulsation frequency. In this study, we cover a broad range of  \mbox{$\dot{\varepsilon}_{\mathrm{{0,set}}}=\unit[0.1-100]{s^{-1}}$} and \mbox{$T=\unit[1-50]{s}$}. Note that we impose a unidirectional pulsatile flow, where the extension rates oscillate with an amplitude of \mbox{$\dot{\varepsilon}_{\mathrm{{0,set}}}$} around constant background flow of \mbox{$\dot{\varepsilon}_{\mathrm{{0,set}}}$}. The resulting extension rate profiles (as commanded to the pumps) along the $x$ and $y$ axes of the OSCER device are shown in Fig.~\ref{FIG_Setup}(b). The constant background flow is maintained for several seconds to stabilize the flow before the superimposed oscillatory modulation begins. We use a multi-function DAQ device (USB-6009, National Instruments, TX) to synchronize the pumps and velocimetry system or polarization camera with a global trigger signal at the beginning of the pulsation cycle.

\subsection{Microparticle image velocimetry}
Applying time-dependent flows in microfluidic systems can lead to significant deviations between the set input signal (\mbox{$\dot{\varepsilon}_{\mathrm{set}}(t)$}) and the actual extension rate profile within the microfluidic chip.\cite{Recktenwald2021b, Recktenwald2025} Therefore, for each imposed oscillatory extension rate profile, we measure the flow field in the OSCER using micro-particle image velocimetry ($\upmu$-PIV, TSI Inc., MN).\cite{Wereley2010, Wereley2019} For this, the OSCER device is mounted on an inverted microscope (Eclipse Ti, Nikon, NY) equipped with a \mbox{$4\times$} air objective (PlanFluor, Nikon, NY) with a numerical aperture of \mbox{$NA=0.13$}. The fluids are seeded with \mbox{$\unit[0.02]{wt.\%}$} of \mbox{$\unit[2]{\upmu m}$} red fluorescent tracer particles (Fluor-Max, Thermo Scientific, Germany) with excitation and emission wavelengths of \mbox{$\unit[542]{nm}$} and \mbox{$\unit[612]{nm}$}, respectively. The tracer particles are excited with a dual-pulsed Nd:YLF laser (\mbox{$\unit[527]{nm}$}) using a volumetric illumination technique. The relative depth over which the tracer particles contribute to the velocity field measurement is \mbox{$\delta_\mathrm{z}\approx0.16 H$}.\cite{Meinhart2000} The flow in the \mbox{$xy$} midplane of the OSCER geometry is recorded using a high-speed camera (Phantom MIRO, Vision Research, Canada), which operates in frame-straddling mode, synchronized with the laser. The frame rate of the camera and the separation of laser pulses are adjusted based on the applied flow rate to achieve an average particle displacement of approximately 4 pixels between consecutive images.

For steady flow measurements, 50 image pairs are recorded and ensemble-averaged over the image sequence. For measurements under pulsatile LAOE, the camera frame rate is set to capture at least 125 image pairs per cycle. At least two full oscillation cycles are recorded per PIV measurement. PIV analysis (TSI Insight 4G, TSI Inc., MN) is performed to obtain the velocity components \mbox{$u$} and \mbox{$v$} in the \mbox{${x}$} and \mbox{${y}$} directions, respectively. Subsequent analysis uses a custom \textsc{MATLAB} (R2024a, The MathWorks, MA) algorithm.

We fit the extension rate \mbox{$\dot{\varepsilon}_{\mathrm{x}}(t)$} along the extension axis measured by PIV with a sinusoidal function, \mbox{$\dot{\varepsilon}_{\mathrm{fit}}(t)=\dot{\varepsilon}_{\mathrm{off}}+\dot{\varepsilon}_{0} \sin(2\pi\, t/T+\varphi)$}. From this fit, the offset \mbox{$\dot{\varepsilon}_{\mathrm{off}}$}, the amplitude \mbox{$\dot{\varepsilon}_{0}$} of the measured signal, and the phase shift \mbox{$\varphi$} between the measured strain rate profile and the set signal are extracted. Hence, the maximum of the temporal strain rate profile is \mbox{$\dot{\varepsilon}_{\mathrm{max}}=\dot{\varepsilon}_{0}+ \dot{\varepsilon}_{\mathrm{off}} $}, which is used for data normalization.

\subsection{Birefringence imaging}
Birefringence imaging is performed using a high-speed polarization camera  (CRYSTA PI-1P, Photron Ltd, Japan) fitted with a \mbox{$10\times$} air objective (PlanFluor, Nikon, NY) with a numerical aperture of \mbox{$NA=0.3$}, and monochromatic light with a wavelength of \mbox{$\unit[520]{nm}$}. We measure the retardance \mbox{$R$} to probe the extent of fluid anisotropy due to the rod alignment during flow and convert it to birefringence as \mbox{$\Delta n = R/(2H)$}. A background value of \mbox{$R\approx\unit[0.5]{nm}$} was determined for the test fluid at rest as the lower detection limit of the device. This noise level is subtracted from all experimental retardation data. The upper detection limit of the device is \mbox{$R\approx\unit[130]{nm}$}, which is not exceeded in our experiments.

Under time-dependent flow conditions, a frame rate between \mbox{$\unit[60-250]{fps}$} is used based on the set extension rate profile to capture at least 250 images per cycle during at least two full oscillation cycles. This data is related to the measured strain rate profile determined via the PIV. We spatially average the birefringence \mbox{$\langle\Delta n \rangle$} and the orientation angle \mbox{$\langle\theta \rangle$} around the central stagnation point along the stretching axis (\mbox{$\lvert x\rvert\leq\unit[0.2]{mm}$} and \mbox{$\lvert y\rvert\leq\unit[0.02]{mm}$}) in each image acquired by the polarization camera.

\subsection{Simulations}\label{sec_methodsim}
We consider a dilute suspension of inertialess rods with a finite aspect ratio, homogeneously dispersed in a Newtonian medium. The adjective dilute implies that each particle is hydrodynamically independent. Each particle has an aspect ratio $r$ and its orientation is defined by a unit vector $\textbf{p}$ parallel to the axis of revolution of the spheroid. In these conditions, the time change of the single spheroid orientation follows the Jeffery's equation:\cite{jeffery1922motion}

\begin{equation}
\dot{\textbf{p}} = \bm{\Omega}\cdot\textbf{p}+ \lambda \left(\textbf{E}\cdot\textbf{p} -\textbf{E}:\textbf{p}\textbf{p}\textbf{p}\right) ,
\label{pdot} 
\end{equation}

\noindent where $\textbf{E}$ is the rate of strain tensor and $\bm{\Omega}$ the vorticity tensor. $\lambda$ represents the spheroid form factor, and it is defined as the ratio $\left(r^{2}-1/r^{2}+1\right)$.  
To describe the evolution of the orientation of a suspension, it is necessary to employ the orientation distribution function $\psi\left(\textbf{p}, t\right)$ defined such that the probability to find a particle oriented within a solid angle $d\textbf{p}$ of $\textbf{p}$ is $\psi d\textbf{p}$. The time evolution of $\psi$ is governed by the following equation: \cite{leal1972rheology}

\begin{equation}
\frac{\partial \psi}{\partial t} + \nabla \cdot \left(\dot{\textbf{p}} \psi - D_r \nabla \psi\right) = 0 .
  \label{psi} 
\end{equation}

\noindent %where $D_r$ is the rotational Brownian diffusion coefficient. 
The orientation distribution evolves because of the torque induced by the external flow field and Brownian motion. 
From the orientation distribution, it is possible to calculate the second-order orientation tensor defined as:
\begin{equation}
 \textbf{a}_{2} = \int \textbf{p}\textbf{p} \psi d\textbf{p} .
  \label{a2} 
\end{equation}
This tensor contains average information on the overall orientation state of the system. Following Fuller (1995),\cite{fuller1995optical} we can express the average anisotropy factor with respect to the flow direction as a function of the orientation tensor as:
\begin{equation}
AF = \sqrt{a_{11}^2-a_{22}^2+4a_{12}^2} ,
  \label{AF} 
\end{equation}

\noindent where $a_{ij}$ are the components of ${\textbf{a}_{2}}$. 
We solve numerically Eq.~\eqref{psi} since no simple analytical solution is possible. We employ a finite volume method as previously performed by Ferec \textit{et al.} \cite{ferec2008numerical, natale2016modeling}  (see Ferec \textit{et al.} \cite{ferec2008numerical} for more details about the mesh and the treatment of the periodic boundary conditions).  A central scheme is implemented to interpolate properties between nodes. Once the orientation distribution is numerically solved, the orientation tensor is evaluated, and the anisotropy factor is consequently obtained. 
In the model, two input parameters are required: $\lambda$ and $D_r$.  These two parameters can be obtained experimentally from AFM particle characterization (Fig.~\ref{FIG_Dist}). For a monodisperse system, the average values of $\lambda$ and $D_r$ (or $\overline{D}_r$) are obtained using the CNC number- and volume-weighted average lengths. 

In order to account for polydispersity, we defined subsets of the suspensions and calculated (for each of the subsets) the average $\overline{D}_r$ starting from the AFM characterization as shown in Fig.~\ref{FIG_Dist}. Since the system is dilute, each subset of particles can be simulated independently from the rest of the subsets. Hence, the $N_{i}$ particles in the subset $i$ have an average rotational diffusion $\overline{D}_{r,i}$ (calculated according eq. \eqref{eq_Dr}), an average form factor $\overline{\lambda}_i$ and average length $\overline{L}_i$. By solving eq. \eqref{psi} with $\overline{D}_{r,i}$ and $\overline{\lambda}_i$, we can calculate the evolution of average $AF$ for the subset $i$, $AF_i$. Once all the subsets are solved, we calculate the following averages:

\begin{equation}
A_{0} = \frac{\sum AF_{i}N_{i} }{\sum N_{i}} ,
  \label{AF_ave} 
\end{equation}

\begin{equation}
A_{1} = \frac{\sum AF_{i}N_{i}\overline{L}_{i} }{\sum N_{i} \overline{L}_{i}} ,
  \label{AF_ave_1} 
\end{equation} 

\begin{equation}
A_{2} = \frac{\sum AF_{i}N_{i}\overline{L}_{i}^2 }{\sum N_{i} \overline{L}_{i}^2} ,
  \label{AF_ave_2} 
\end{equation} 

\begin{equation}
A_{3} = \frac{\sum AF_{i}N_{i}\overline{L}_{i}^3 }{\sum N_{i} \overline{L}_{i}^3} .
  \label{AF_ave_3} 
\end{equation} 

These averages consider the impact on birefringence given by the number of particles $A_0$, the size of the particle $A_1$, the area that the particles span $A_2$,  and the volume that the particles $A_3$ cover during motion. A similar approach was employed to explain the effect of polydispersity in graphene suspensions under simple shear flow, where it was found that the dichroic response was strongly influenced by the area spanned by each particle during rotation.\cite{reddy2018rheo}

\section{Results and Discussion}
First, we probe the flow behavior and birefringence of the CNC test fluids under steady flow conditions in Sec.~\ref{sec_steady} to validate the homogeneity of the flow field with the covered strain rate range. The experimental results are compared to simulations of monodisperse and polydisperse systems. Second, we employ pulsatile flow conditions and test the dynamic behavior of the rods under LAOE in Sec.~\ref{sec_LAOE}, focusing on the temporal evolution of the birefringence. 

Note that in all experiments, we define the Reynolds number, describing the ratio of inertial to viscous forces during flow, as \mbox{$Re = \rho U D_{\mathrm{h}} / \eta$}, where the fluid density, \mbox{$\rho=\unit[1215]{kg/m^3}$}, the hydraulic diameter of the rectangular inlet and outlet channels \mbox{$D_{\mathrm{h}} = 4WH/(W+H)=0.364~\text{mm}$}, and fluid viscosity \mbox{$\eta$}. Based on the maximum velocity reached in the experiments, the maximum Reynolds number reached in this study is \mbox{$Re \approx 0.23$}.

\subsection{Steady flow conditions}\label{sec_steady}
\subsubsection{Flow field characterization}
\begin{figure}[ht]
\centering
\includegraphics[width=0.45\textwidth]{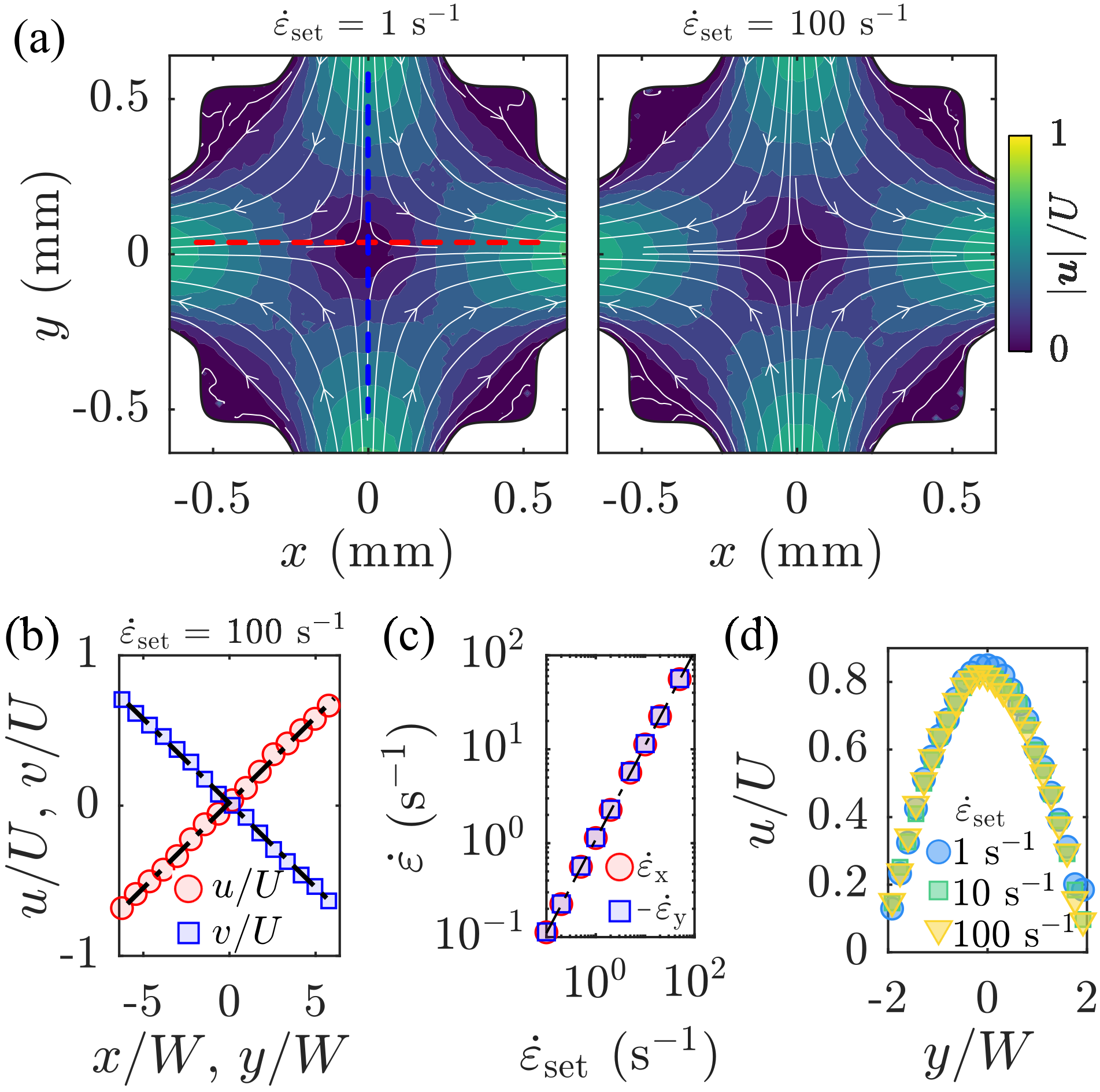}
\caption{Flow velocimetry obtained with the \mbox{$\unit[0.1]{\%}$} CNC dispersion under steady flow conditions. (a) Normalized velocity field with superimposed streamlines at \mbox{$\dot{\varepsilon}_{\mathrm{set}}=\unit[1]{s^{-1}}$} (left) and \mbox{$\dot{\varepsilon}_{\mathrm{set}}=\unit[100]{s^{-1}}$} (right). (b) Normalized velocity components along the ${x}$ and ${y}$ directions representatively shown for \mbox{$\dot{\varepsilon}_{\mathrm{set}}=\unit[100]{s^{-1}}$}. Black dashed lines represent linear fits over \mbox{$\lvert y/W \rvert \leq 6$} and \mbox{$\lvert x/W \rvert \leq 6$}, which are used to calculate the extension rates \mbox{$\dot{\varepsilon}_{\mathrm{x}}$} and \mbox{$\dot{\varepsilon}_{\mathrm{y}}$}, respectively. (c) Extension rates along the ${x}$ and ${y}$ directions as a function of the set elongation rate. (d) Streamwise velocity profiles normalized by the centerline velocity, measured across an outlet \mbox{$\unit[7]{mm}$} downstream of the stagnation point at various imposed \mbox{$\dot{\varepsilon}_{\mathrm{set}}$}.}
\label{FIG_SteadyPIV}
\end{figure}

Under conditions of steady extensional flow, the flow field measured for all of the test fluids remains Newtonian-like over the entire range of investigated strain rates (i.e., \mbox{$0.1 \leq \dot{\varepsilon}_{\mathrm{set}} \leq \unit[100]{s^{-1}}$}, \mbox{$\text{Re}<0.23$}), maintaining symmetry about the central stagnation point and the principal flow axes. This is exemplified in Fig.~\ref{FIG_SteadyPIV}(a) for the \mbox{$\unit[0.1]{\%}$} CNC sample at two representative strain rates. The velocity components \mbox{$v$} (in ${y}$ direction) and \mbox{$u$} (in ${x}$ direction) are extracted along their respective axes, indicated by the dashed lines in Fig.~\ref{FIG_SteadyPIV}(a). Over the range of flow rates studied, \mbox{$u$} increases, and \mbox{$v$} decreases linearly along their respective axes (representatively shown for \mbox{$\dot{\varepsilon}_{\mathrm{set}}=\unit[100]{s^{-1}}$} in Fig.~\ref{FIG_SteadyPIV}(b)). The average strain rate in ${x}$ direction (\mbox{$\dot{\varepsilon}_{\mathrm{x}}=\partial u/\partial x$}) and ${y}$ direction (\mbox{$\dot{\varepsilon}_{\mathrm{y}}=\partial v/\partial y$}) are calculated by averaging the velocity gradients over the spatial domain \mbox{$\lvert y/W \rvert \leq 6$} and \mbox{$\lvert x/W \rvert \leq 6$} (indicated by the black dashed lines in Fig.~\ref{FIG_SteadyPIV}(b)). 

The magnitudes of the average extension rates along the compression and elongation axes are the same and increase linearly as a function of the set extension rate \mbox{$\dot{\varepsilon}_{\mathrm{set}}$} (Fig.~\ref{FIG_SteadyPIV}(c)). 

Across all investigated strain rates, we observe parabolic profiles of the streamwise velocity across the outlet and inlet channels (Fig.~\ref{FIG_SteadyPIV}(d)) without the emergence of any flow modifications, which have been reported for flows of polymeric systems even at $c<c^*$.\cite{Dunlap1987, Haward2012d, Haward2023b, Recktenwald2025} Taken together, the fact that \mbox{$\dot{\varepsilon}_{\mathrm{x}}=-\dot{\varepsilon}_{\mathrm{y}} = \dot{\varepsilon}_{set}$} and the absence of flow modification at both the outlet and inlet imply that the CNC dispersions behave analogously to a Newtonian fluid, consistent with previous observations for rigid colloidal rods at \mbox{$c \lesssim c^*$}.\cite{Calabrese2021,Calabrese2024b}

Based on the homogeneity of the flow field during planar extension for the used CNC systems in this study, we define  \mbox{$\dot{\varepsilon}\equiv\dot{\varepsilon}_{\mathrm{x}}$} and \mbox{$\dot{\varepsilon}(t)\equiv\dot{\varepsilon}_{\mathrm{x}}(t)$} as characteristic strain rate imposed on the fluid along the stretching axis.

\subsubsection{Birefringence}
\begin{figure}[ht!]
\centering
\includegraphics[width=0.45\textwidth]{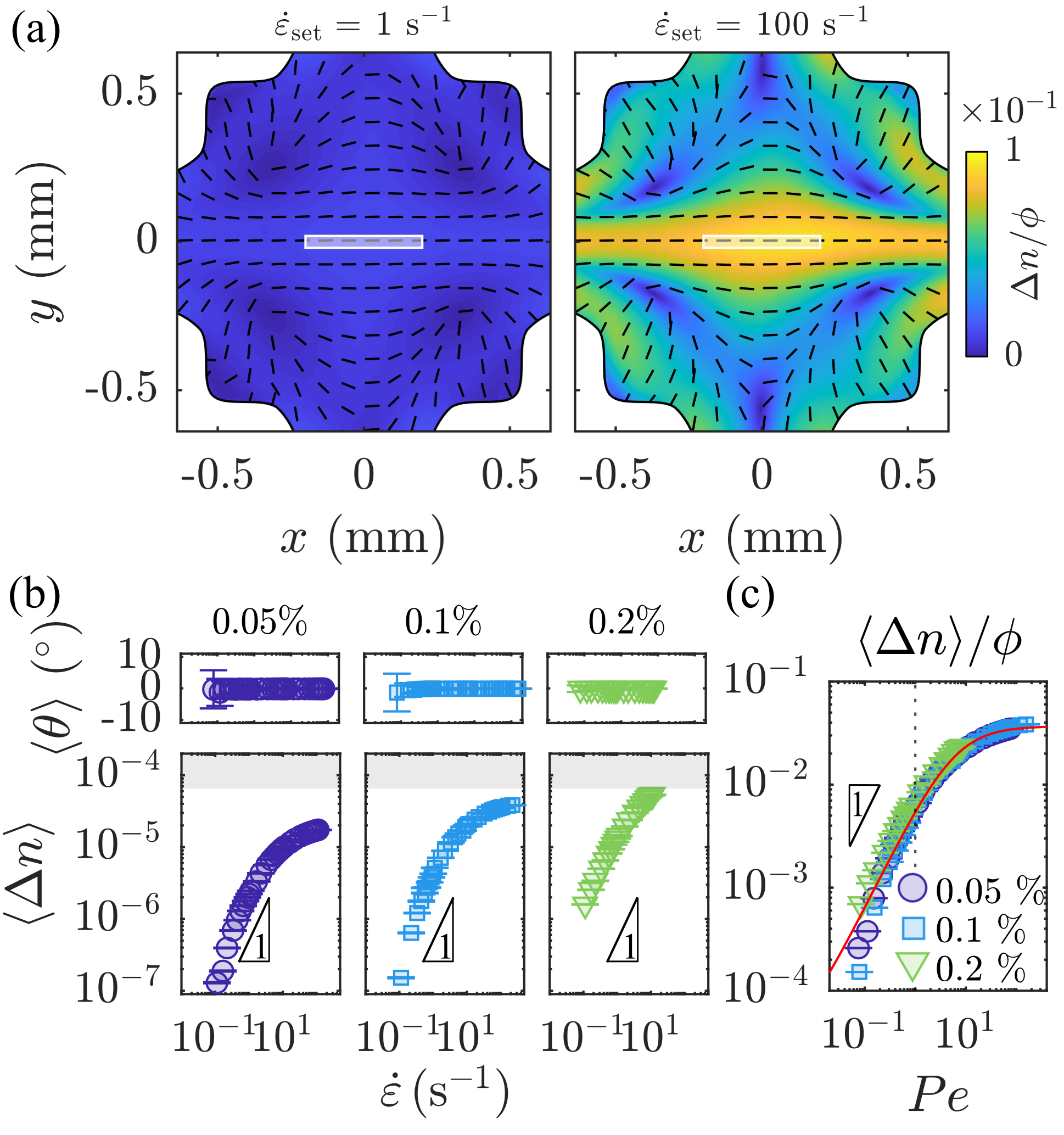}
\caption{Birefringence of the CNC dispersions under steady flow conditions. (a) Time-averaged birefringence normalized by the CNC concentration for the same conditions as in Fig.~\ref{FIG_SteadyPIV}(a). The orientation angle \mbox{$\theta$} is indicated by the solid segments. The white boxes around the central stagnation point in (a) correspond to regions of \mbox{$\lvert x\rvert\leq\unit[0.2]{mm}$} and \mbox{$\lvert y\rvert\leq\unit[0.02]{mm}$}, where the spatially-averaged data is calculated. (b) Spatially-averaged orientation angle \mbox{$\langle \theta \rangle$}(top) and birefringence \mbox{$\langle\Delta n \rangle$} (bottom) for the three samples as a function of the extension rate \mbox{$\dot{\varepsilon}$}. Gray areas in (b) indicate the polarization camera's maximum retardance limit. (c) Spatial averaged birefringence \mbox{$\langle\Delta n \rangle$} scaled with the CNC volume fraction \mbox{$\phi$} as a function of P\'eclet number \mbox{$Pe$}, using \mbox{$\overline {D}_r = \unit[1.44]{s^{-1}}$}. The red solid line shows a fit to the data according to Eq.~\ref{eq_steadyFIB}.}
\label{FIG_SteadyFIB}
\end{figure}

Under steady flow conditions, we observe the emergence of a relatively large area of birefringence around the central stagnation point in the OSCER device as the extension rate is increased (Fig.~\ref{FIG_SteadyFIB}(a)). This relatively large region of strong birefringence around the stagnation point can be attributed to the small fluid strain required for the rods to attain a preferential orientation. This behavior contrasts sharply with that of flexible polymer solutions, which exhibit a highly localized birefringent strand along the extension axis, since polymer chains must experience a substantially larger fluid strain before undergoing stretching and backbone alignment.

In the microfluidic OSCER device, the CNC aligns perpendicularly to the flow direction along the compressional ${y}$-axis due to the deceleration of the fluid, as indicated by the solid black segments in Fig.~\ref{FIG_SteadyFIB}(a). In contrast, the CNC aligns parallel to the flow direction along the elongation axis in the ${x}$ direction. In previous studies, similar orientation patterns have been reported for CNC systems and other anisotropic particles.\cite{Kiriya2012, Trebbin2013a, Corona2018, Calabrese2021} We spatially average the orientation angle \mbox{$\langle \theta \rangle$} in the central part of the OSCER around the stagnation point (white boxes in Fig.~\ref{FIG_SteadyFIB}(a)) to quantify the CNC alignment along the extension axis. We observe alignment in the flow direction with \mbox{$\langle \theta \rangle \approx \unit[0]{^{\circ}}$} over the entire range of tested \mbox{$\dot{\varepsilon}$} (Fig.~\ref{FIG_SteadyFIB}(b), top), consistent with a previous report.\cite{Calabrese2021}

The average birefringence \mbox{$\langle\Delta n \rangle$}, also spatially-averaged in the central part of the OSCER around the stagnation point (white boxes in Fig.~\ref{FIG_SteadyFIB}(a)), increases with increasing extension rate \mbox{$\dot{\varepsilon}$} for all investigated samples (Fig.~\ref{FIG_SteadyFIB}(b), bottom). At low strain rates, the birefringence increases almost linearly with a slope of 1 for \mbox{$\dot{\varepsilon} \lesssim \unit[1]{s^{-1}}$}. Increasing the strain rate results in a progressive saturation of the birefringence. For \mbox{$\dot{\varepsilon} \gtrsim \unit[10]{s^{-1}}$}, the birefringence approaches a plateau for \mbox{$\unit[0.05]{\%}$} and \mbox{$\unit[0.1]{\%}$} CNC (Fig.~\ref{FIG_SteadyFIB}(b), bottom). For the \mbox{$\unit[0.2]{\%}$} CNC suspension, the birefringence at the saturation plateau lies beyond the maximum retardance limit of the polarization camera.

Since the birefringence intensity is proportional to the number of aligned rods in the optical path, we normalize the averaged birefringence data \mbox{$\langle\Delta n \rangle$} of Fig.~\ref{FIG_SteadyFIB}(b) by the corresponding CNC volume fraction \mbox{$\phi$} of the samples (\mbox{$\phi=\mathrm{vol}\%/100$}). The resulting \mbox{$\langle\Delta n \rangle/\phi$} curves show an overlap as a function of the P\'eclet number, here defined as $Pe=\dot\varepsilon/\overline {D}_r$, with \mbox{$\overline {D}_r = \unit[1.44]{s^{-1}}$} (Fig.~\ref{FIG_SteadyFIB}(c)). This data collapse highlights the linear increase with a slope of 1 for \mbox{$Pe \lesssim 1$}, the progressive saturation, and ultimately the plateau in birefringence at large \mbox{$Pe \gtrsim 10$}. Since $D_r$ is constant and concentration-independent in the dilute regime (see Eq.~\ref{eq_Dr}), the good collapse of \mbox{$\langle \Delta n \rangle / \phi$} vs.\ $Pe$ for the CNC at different concentrations confirms that the rods are effectively dilute and non-interacting at the tested concentrations.

This trend in birefringence as a function of $Pe$ can be described, similarly to Santos \textit{et al.}\cite{santos2023flow}, by the following empirical relation:

\begin{equation}
\frac{\Delta n}{\phi}=\frac{\Delta n_{\mathrm{max}}}{\phi}\left[ 1- \frac{1}{1+\left(\frac{Pe}{a}\right)^b}\right].
    \label{eq_steadyFIB} 
\end{equation}

By fitting Eq.~\ref{eq_steadyFIB} to the experimental data shown in Fig.~\ref{FIG_SteadyFIB}(c), we obtain \mbox{$\Delta n_{\mathrm{max}} / \phi \approx 3.8 \times 10^{-2}$} as the plateau value of the normalized birefringence at high $Pe$, \mbox{$a = 6$} as the inflection point of the sigmoidal curve, and \mbox{$b \approx 1$} as the slope parameter. The value of \mbox{$\Delta n_{\mathrm{max}} / \phi = 3.8 \times 10^{-2}$}, often referred to as the intrinsic birefringence, is in good agreement with those reported for cellulose-based structures.\cite{uetani2019estimation}

\subsubsection{Simulations}\label{sec_SIMsteady}
\begin{figure}[ht]
\centering
\includegraphics[width=0.45\textwidth]{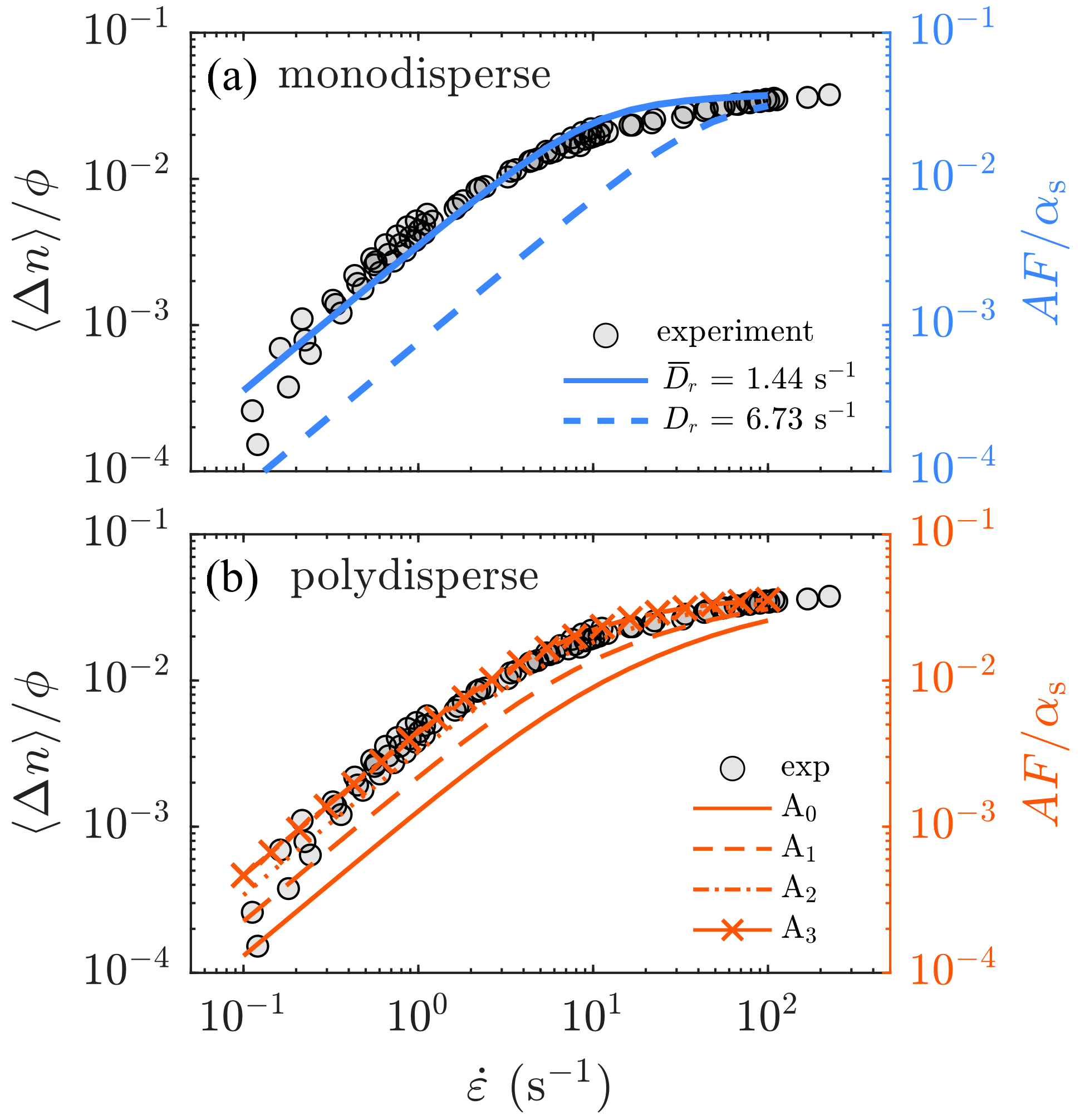}
\caption{Simulation results for (a) a monodisperse system and (b) polydisperse systems showing the anisotropy factor \mbox{$AF$} scaled by a constant factor \mbox{$\alpha_{\mathrm{s}}$}. Experimental data of normalized birefringence are superimposed. In (a), two simulations for different diffusion coefficients are shown corresponding to calculations of Eq.\ref{eq_Dr} using the volume-averaged contour length \mbox{${\langle \overline{l_{\mathrm{c}}} \rangle}=\unit[470]{nm}$} (\mbox{$\overline {D}_r = \unit[1.44]{s^{-1}}$}) and the number-averaged contour length \mbox{$\langle l_{\mathrm{c}}\rangle=\unit[260]{nm}$} (\mbox{$D_r=\unit[6.73]{s^{-1}}$}). (b) Simulations for different averages \mbox{$A_{0-3}$} using \mbox{$N=10$} bins based on Fig.~\ref{FIG_Dist}(b). }
\label{FIG_SteadySIM} 
\end{figure}

Figure~\ref{FIG_SteadySIM} compares the experimental birefringence data \mbox{$\langle\Delta n \rangle/\phi$} (see Fig.~\ref{FIG_SteadyFIB}(c)) with numerical simulations for monodisperse (Fig.~\ref{FIG_SteadySIM}(a)) and polydisperse systems (Fig.~\ref{FIG_SteadySIM}(b)). For the monodisperse case, two simulations were performed using \mbox{$\overline {D}_r=\unit[1.44]{s^{-1}}$} and \mbox{$D_r=\unit[6.73]{s^{-1}}$}, based on the volume-averaged contour length \mbox{${\langle \overline{l_{\mathrm{c}}}\rangle}=\unit[470]{nm}$} and the number-averaged contour length \mbox{$\langle l_{\mathrm{c}}\rangle=\unit[260]{nm}$}, respectively. The degree of CNC alignment in the simulation is quantified by the anisotropy factor \mbox{$AF$}, which ranges from 0 to 1. We scale $AF$ by $\alpha_{\mathrm{s}}=1/({\Delta n_{\mathrm{max}} / \phi} )\approx  26$, \textit{i.e.}, as $AF/\alpha_{\mathrm{s}}$, to enable a direct comparison with the experimental birefringence data. This scaling is applied uniformly to all numerical data for both monodisperse and polydisperse systems.

In both cases, the anisotropy factor \mbox{$AF$} increases with strain rate \mbox{$\dot{\varepsilon}$}, similarly to the experimental birefringence measurements. The main difference between the two simulations is that, for the monodisperse simulations with \mbox{$D_r=\unit[6.73]{s^{-1}}$}, $AF$ curve is shifted to higher extension rates compared to the $AF$ obtained using \mbox{$\overline{D}_r=\unit[1.44]{s^{-1}}$}.

In the polydisperse simulations, the different averages \mbox{$A_{0}$} to \mbox{$A_{3}$} for \mbox{$AF$} are calculated using \mbox{$N=10$} bins based on the distribution of \mbox{$D_r$} (Fig.~\ref{FIG_Dist}(b)), as described in Sec.~\ref{sec_methodsim}.  Importantly, as the weight function shifts from \mbox{$A_{0}$} to \mbox{$A_{3}$}, giving higher weight to longer rods, the average contour length increases. For all polydisperse simulations, the average \mbox{$AF$} increases with strain rate and eventually reaches a plateau at high strain rates. Scaling the numerical \mbox{$AF$} data with \mbox{$\alpha_{\mathrm{s}}=26$} yields excellent agreement of \mbox{$A_{3}$} with the experimental data across the entire strain rate range. The other averages also exhibit a qualitatively similar increase in \mbox{$AF$}. However, a progressive shift of \mbox{$AF$} toward higher extension rates is observed as the weight function shifts to lower powers of the rod length, consistent with the expected increase in the effective value of $D_r$.

Overall, the birefringence experiments under steady flow conditions show good agreement with both the monodisperse simulation, based on the volume-averaged contour length (\mbox{$\overline{D}_r=\unit[1.44]{s^{-1}}$}), and the polydisperse simulation with \mbox{$A_{3}$}.

\subsection{Rod dynamics under LAOE}\label{sec_LAOE}
\subsubsection{Time-dependent birefringence}\label{sec_FIBtime}
We next investigate the time-dependent response of the CNC systems to LAOE over a broad range of set strain rate amplitudes \mbox{$\dot{\varepsilon}_{\mathrm{0,set}}$} and oscillation periods \mbox{$T$}. 

For time-dependent measurements, we define \mbox{$Pe_0= \dot{\varepsilon}_{0} / \overline{D}_r$} using the extension rate amplitude \mbox{$\dot{\varepsilon}_{0}$} determined from the time-dependent strain rate in the PIV measurements. Moreover, we define the Deborah number \mbox{$De =f/\overline{D}_r= (\overline{D}_r \, T)^{-1}$}, which relates the diffusion timescale to the pulsation period in the time-dependent measurements.

\begin{figure*}[ht]
\centering
\includegraphics[width=\textwidth]{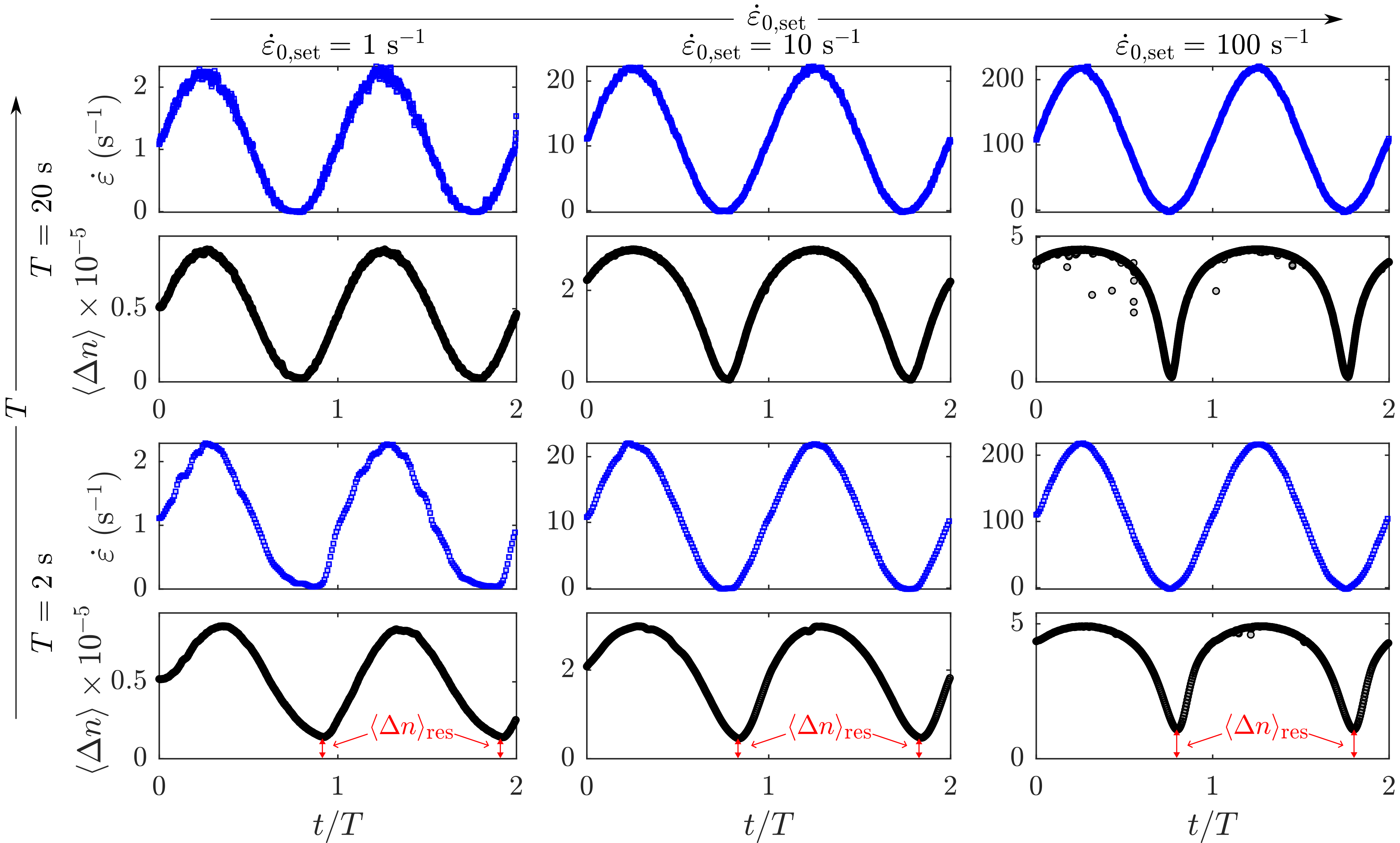}
\caption{Time-dependent strain rate \mbox{$\dot{\varepsilon}(t)$} (top sub-panels) and average birefringence \mbox{$\langle\Delta n \rangle(t)$} (bottom sub-panels) in the experiments under pulsatile LAOE. Example data is shown for the for \mbox{$\unit[0.1]{\%}$} sample over two normalized pulsation cycles \mbox{$t/T$} at various \mbox{$\dot{\varepsilon}_{\mathrm{{0,set}}}$} for \mbox{$T=\unit[2]{s}$} and \mbox{$T=\unit[20]{s}$}. The residual birefringence \mbox{$\langle\Delta n \rangle_{\mathrm{res}}$} that does not decay to zero during the cycle at higher frequencies is indicated in the bottom panels for the measurements at \mbox{$T=\unit[2]{s}$}.}
\label{FIG_PulseEXP}
\end{figure*}

Figure~\ref{FIG_PulseEXP} presents representative results in terms of the temporal average birefringence \mbox{$\langle\Delta n \rangle(t)$} for the \mbox{$\unit[0.1]{\%}$} sample. At low strain rates and large period durations, \textit{e.g.}, \mbox{$\dot{\varepsilon}_{\mathrm{{0,set}}}=\unit[1]{s^{-1}}$} and \mbox{$T=\unit[20]{s}$} in Fig.~\ref{FIG_PulseEXP}, the measured birefringence \mbox{$\langle\Delta n \rangle(t)$} follows the sinusoidal waveform of the imposed strain rate \mbox{$\dot{\varepsilon}(t)$} along the extension axis. However, increasing the set strain rate amplitude \mbox{$\dot{\varepsilon}_{\mathrm{{0,set}}}$} while keeping the period duration fixed at \mbox{$T=\unit[20]{s}$} results in a progressive deviation from the sinusoidal waveform of the birefringence response. Although the strain rate still exhibits a sine, we observe a flattening of the birefringence profile \mbox{$\langle\Delta n \rangle(t)$} at large \mbox{$\dot{\varepsilon}(t)$} during the cycle, \textit{e.g.}, at \mbox{$\dot{\varepsilon}_{\mathrm{{0,set}}}=\unit[10]{s^{-1}}$} and \mbox{$T=\unit[20]{s}$}. Increasing the extension rate amplitude further enhances the magnitude of this phenomenon and results in a saturation of \mbox{$\langle\Delta n \rangle(t)$} during the ascending phase of the sine. Note that this saturation during the cycle occurs below the polarization camera's maximum retardance detection limit (\mbox{$\equiv \Delta n \approx\unit[7\times]{10^{-5}}$}). During the LAOE cycle at large period durations, the birefringence decays back to zero at \mbox{$\dot{\varepsilon}(t)=0$} and the  \mbox{$\langle\Delta n \rangle(t)$} response at the descending and ascending phases of the sine appear to be symmetrical with respect to the minimum birefringence during the cycle.

Upon increasing the pulsation frequency and applying a low strain rate amplitude, \textit{e.g.}, \mbox{$\dot{\varepsilon}_{\mathrm{{0,set}}}=\unit[1]{s^{-1}}$} and \mbox{$T=\unit[2]{s}$} in Fig.~\ref{FIG_PulseEXP}, the birefringence response follows again the imposed strain rate profile. Note that the deviation between the true extensional rate in the OSCER \mbox{$\dot{\varepsilon}(t)$} and the pure sinusoidal set strain rate at the pump \mbox{$\dot{\varepsilon}_{\mathrm{set}}(t)$} emerges as the syringe pumps approach their operational lower minimum flow rate limit.\cite{Recktenwald2025} Nevertheless, due to our PIV analysis, we can correlate deviations from a sine waveform in the response \mbox{$\langle\Delta n \rangle(t)$} with true input extension rate signal \mbox{$\dot{\varepsilon}(t)$}. Increasing the set strain rate amplitude \mbox{$\dot{\varepsilon}_{\mathrm{{0,set}}}$} results in a saturation of \mbox{$\langle\Delta n \rangle(t)$} as the strain rate increases during the cycle, similar to the observations at \mbox{$T=\unit[20]{s}$}. However, the birefringence response at the descending and ascending phases of the sine appears to become increasingly asymmetric. In contrast to the response at low frequencies (\mbox{$T=\unit[20]{s}$} in Fig.~\ref{FIG_PulseEXP}), we observe that the birefringence does not decay back to zero when \mbox{$\dot{\varepsilon}(t)\rightarrow0$} but exhibits a residual birefringence \mbox{$\langle\Delta n \rangle_{\mathrm{res}}$} during the cycle (see red arrows in the bottom panels of Fig.~\ref{FIG_PulseEXP}).

Based on these observations, we can distinguish between two contrasting regimes, depending on the pulsation period and the relaxation timescale of the rods. For sufficiently long pulsation periods, \mbox{$T \gg 1/\overline{D}_r$}, hence \mbox{$De\ll1$}, the rod dynamics closely resemble those observed under steady-state conditions, as the rods have enough time to adopt an average orientation comparable to that in steady flow during both the ascending and descending parts of the imposed sinusoidal strain rate cycle. Importantly, in this regime, the rods have sufficient time to fully relax and reach an anisotropic distribution at \mbox{$\dot{\varepsilon} < \overline{D}_r$}, with the birefringence signal decaying to the noise floor. In contrast, for relatively short pulsation periods, $T \sim 1/\overline{D}_r$, the pulsation timescale becomes comparable to or faster than the intrinsic rod relaxation timescale. As a consequence, the rod dynamics are highly transient and strongly dependent on their orientational history. This manifests itself as a progressively increasing asymmetry in the birefringence response with a decreasing pulsation period, together with a finite residual alignment persisting even at \mbox{$\dot\varepsilon = 0$}.

\subsubsection{Simulations}\label{sec_FIBtimeSIM}
\begin{figure*}[ht]
\centering
\includegraphics[width=\textwidth]{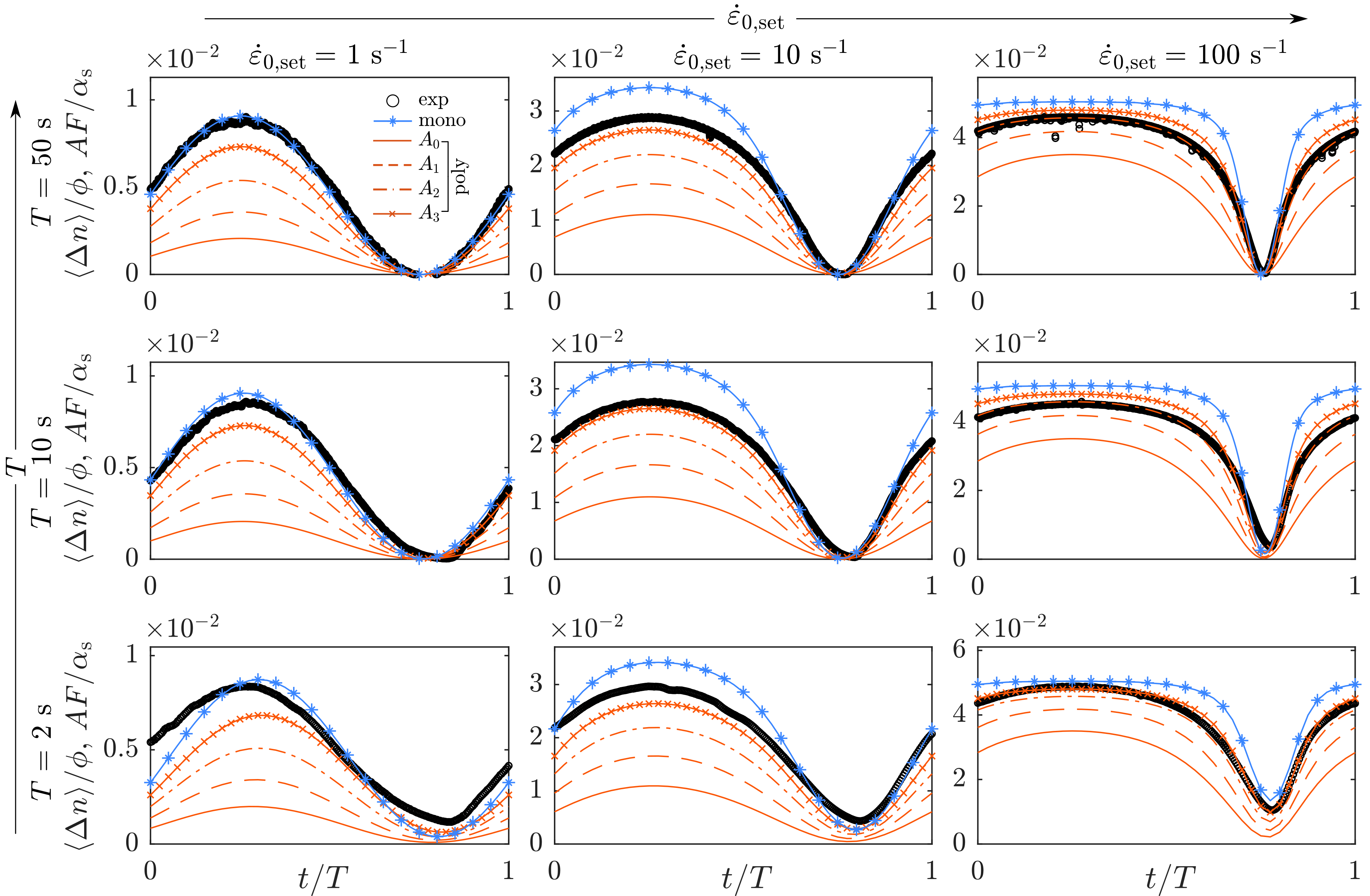}
\caption{Comparison of the numerical simulations and experiments under LAOE. Normalized time-dependent average birefringence signal \mbox{$\langle\Delta n \rangle/\phi$} for the experiment and anisotropy factor  \mbox{$AF/\alpha_{\mathrm{s}}$} for the simulations under pulsatile LAOE. Exemplary experimental data are shown for the \mbox{$\unit[0.1]{\%}$} CNC sample, and the simulations show data for a monodisperse system with \mbox{$\overline{D}_r = \unit[1.44]{s^{-1}}$} and for the polydisperse systems. Data is shown for various \mbox{$T$} and \mbox{$\dot{\varepsilon}_{\mathrm{{0,set}}}$} over one normalized pulsation cycle \mbox{$t/T$}.
}
\label{FIG_PulseSIM}
\end{figure*}

Figure~\ref{FIG_PulseSIM} compares the experimental LAOE results of the \mbox{$\unit[0.1]{\%}$} sample against the corresponding numerical simulations for monodisperse and polydisperse systems. Representative data is shown for various \mbox{$T$} and \mbox{$\dot{\varepsilon}_{\mathrm{{0,set}}}$} over one normalized pulsation cycle \mbox{$t/T$}.

At low strain rate amplitudes and large pulsation periods, \textit{e.g.}, \mbox{$\dot{\varepsilon}_{\mathrm{0,set}}=\unit[1]{s^{-1}}$} and \mbox{$T=\unit[50]{s}$} in Fig.~\ref{FIG_PulseSIM}, the numerical data  \mbox{$AF/\alpha_{\mathrm{s}}$} follows a sinusoidal profile throughout the cycle, similar to the time-dependent average birefringence \mbox{$\langle\Delta n \rangle/\phi$} observed in the experiment. In this case, the monodisperse simulations show excellent agreement with the experimental data. Similar to the steady flow conditions, \mbox{$A_{3}$} provides the best agreement for the polydisperse simulations, while the other averages progressively underestimate the birefringence. This trend among the different averages is observed for most tested combinations of strain rate amplitudes and pulsation periods under LAOE.

Increasing the set strain rate amplitude \mbox{$\dot{\varepsilon}_{\mathrm{0,set}}$} while keeping the period fixed at \mbox{$T=\unit[50]{s}$} (Fig.~\ref{FIG_PulseSIM}) leads to a progressive deviation from the sinusoidal waveform of \mbox{$AF/\alpha_{\mathrm{s}}$} in both monodisperse and polydisperse simulations, in qualitative agreement with the experimental birefringence response. At \mbox{$\dot{\varepsilon}_{\mathrm{0,set}}=\unit[100]{s^{-1}}$}, the polydisperse simulation with \mbox{$A_{3}$} captures the experimental data very well. In contrast, the monodisperse simulation exhibits a faster relaxation and rise of \mbox{$AF/\alpha_{\mathrm{s}}$} during the descending and ascending phases of the cycle compared to the experiment. A qualitatively similar evolution of \mbox{$AF/\alpha_{\mathrm{s}}$} during the sinusoidal strain rate modulation is observed for both monodisperse and polydisperse systems across all simulations with increasing strain rate amplitude.

At \mbox{$T=\unit[2]{s}$}, we also observe a transition from a sinusoidal to a more saturated (\textit{i.e.}, flat-topped) waveform in simulations as the strain rate amplitude is increased. In addition, similar to the experimental birefringence, a distinct residual in the anisotropy factor \mbox{$AF_{\mathrm{res}}/\alpha_{\mathrm{s}}$} appears in all simulations as the temporal strain rate approaches zero during the cycle. The magnitude of \mbox{$AF_{\mathrm{res}}/\alpha_{\mathrm{s}}$} increases from \mbox{$A_0$} to \mbox{$A_3$} in the polydisperse simulations and also rises with increasing strain rate amplitude at a fixed LAOE period.

Overall, the temporal evolution of \mbox{$AF/\alpha_{\mathrm{s}}$} during the LAOE cycle qualitatively matches the experimental observations. Both the saturation at high strain rates and the residual birefringence when the flow temporarily ceases are well captured in the simulations. Among them, the polydisperse simulation with \mbox{$A_3$} shows the best agreement with the experimental data, indicating the important influence of the longer rods on the overall birefringence signal.

\begin{figure}[ht!]
\centering
\includegraphics[width=0.37\textwidth]{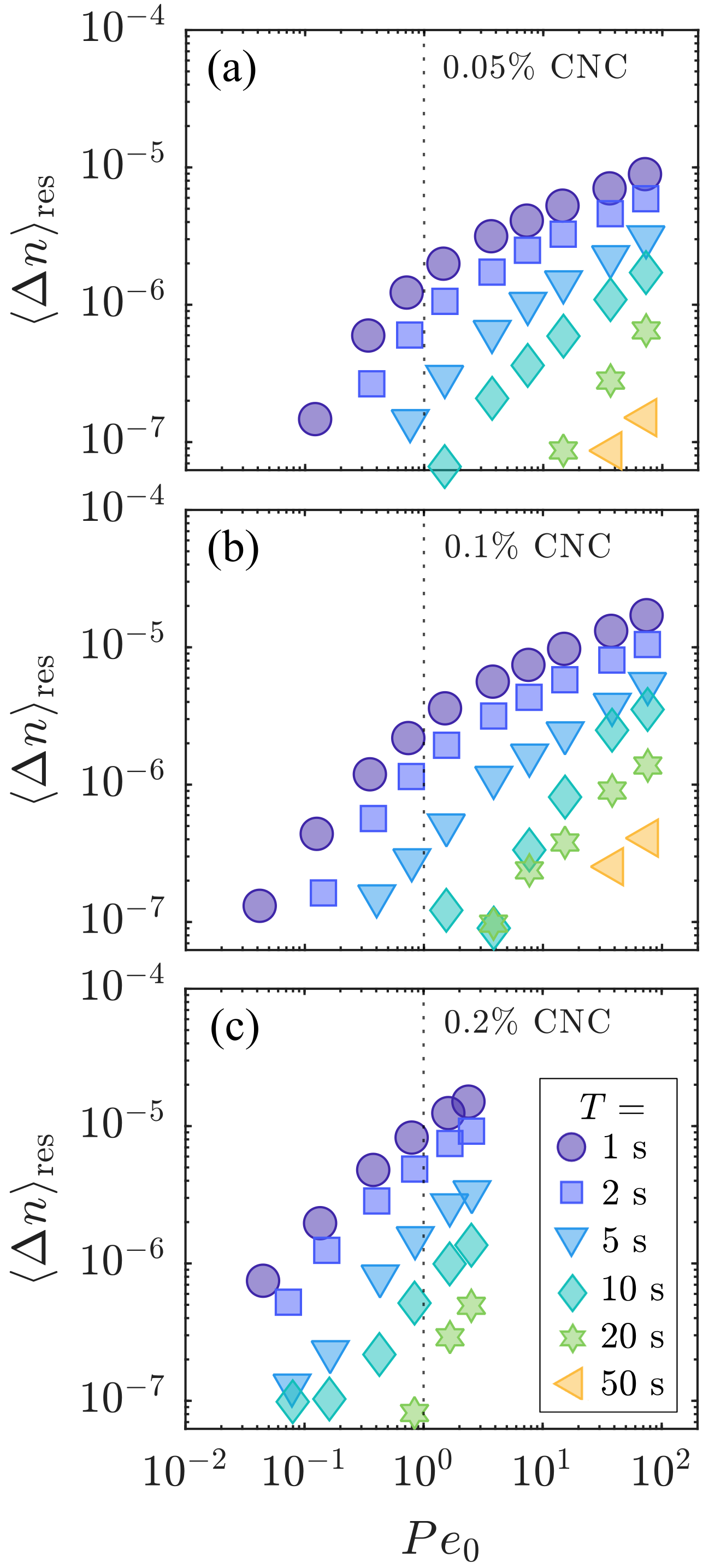}
\caption{Residual birefringence \mbox{$\langle\Delta n \rangle_{\mathrm{res}}$} during pulsatile LAOE. Data is shown as a function of \mbox{$Pe_0$} for the three samples (a-c) and various periods \mbox{$T$}.  
}
\label{FIG_Residual}
\end{figure}

\subsubsection{Residual birefringence during LAOE}\label{sec_FIBres}
As shown in Figs.~\ref{FIG_PulseEXP} and \ref{FIG_PulseSIM}, a residual birefringence \mbox{$\langle\Delta n \rangle_{\mathrm{res}}$} (or \mbox{$AF_{\mathrm{res}}/\alpha_{\mathrm{s}}$} in the simulations) emerges during LAOE as the pulsation frequency is increased. Figure~\ref{FIG_Residual}(a–c) summarizes the experimental observations of \mbox{$\langle\Delta n \rangle_{\mathrm{res}}$} for the three CNC samples as a function of \mbox{$Pe_0= \dot\varepsilon_0 / \overline{D}_r$} at various period durations \mbox{$T$}.

At long period durations, \textit{e.g.}, \mbox{$T=\unit[50]{s}$} in Fig.~\ref{FIG_Residual}(a), \mbox{$\langle\Delta n \rangle_{\mathrm{res}}\approx0$} until \mbox{$Pe_0\approx30$}, where the residual birefringence begins to increase with rising P\'eclet number. As the period duration decreases, \textit{i.e.}, the pulsation frequency increases, the onset of detectable residual birefringence \mbox{$\langle\Delta n \rangle_{\mathrm{res}}$} shifts to lower \mbox{$Pe_0$} values. At \mbox{$T=\unit[1]{s}$}, the birefringence no longer decays to zero during periodic extensional flow for all tested \mbox{$Pe_0\geq4\times10^{-2}$} (Fig.~\ref{FIG_Residual}(a)). A qualitatively similar trend is observed for all CNC systems (Fig.~\ref{FIG_Residual}(b–c)). However, at a fixed period duration and \mbox{$Pe_0$}, the residual birefringence is higher for samples with greater CNC concentration, consistent with the behavior under steady flow conditions (see Fig.~\ref{FIG_SteadyFIB}(b)).

\begin{figure}
\centering
\includegraphics[width=0.43\textwidth]{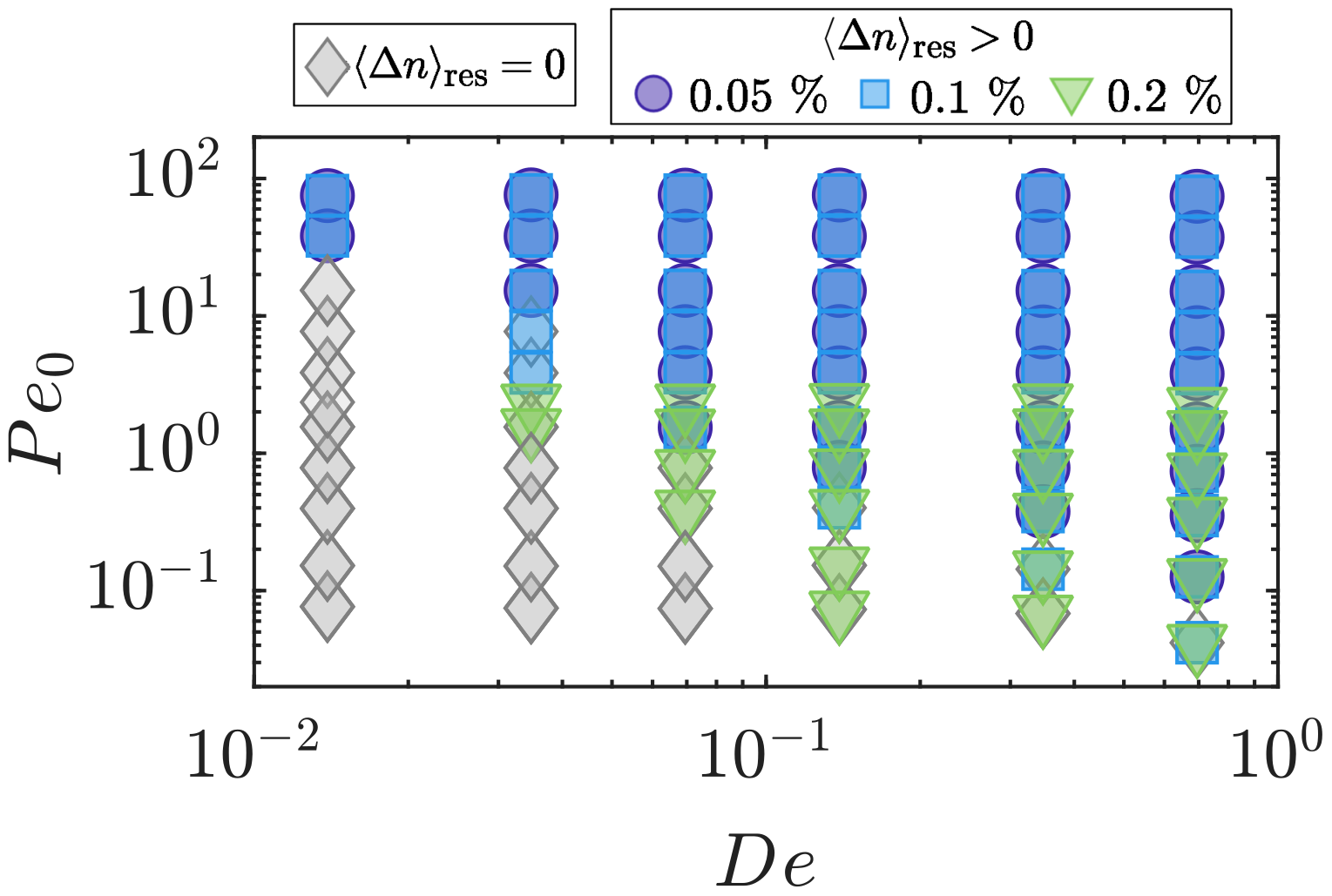}
\caption{Pipkin space representing all experiments for the different CNC systems at different \mbox{$Pe_0$} and \mbox{$De$}. Colored symbols indicate measurements where a residual birefringence is measurable, \textit{i.e.}, larger than the polarization camera's lower detection limit. Gray symbols show measurements where the residual birefringence is within the polarization camera's lower detection limit.}
\label{FIG_Pipkin} 
\end{figure}

To further explore the critical conditions for the emergence of residual birefringence during the LAOE cycle in our experiments, we determine the values of \mbox{$Pe_0$} and \mbox{$De$} at which non-zero residual birefringence, exceeding the noise threshold, first occurs. Figure~\ref{FIG_Pipkin} shows a Pipkin space representation of the experimental results across various \mbox{$Pe_0-De$} combinations. Colored symbols indicate measurements where a measurable residual birefringence is observed, while gray symbols correspond to measurements where the residual birefringence remains within the lower detection limit of the polarization camera.

At low pulsation frequencies, \textit{e.g.}, \mbox{$De<0.02$}, a residual birefringence emerges only above \mbox{$Pe_0\gtrsim20$}. As the LAOE frequency, and thus \mbox{$De$}, increases, the onset of residual birefringence shifts toward lower critical P\'eclet numbers. For \mbox{$De\approx0.05$}, the critical \mbox{$Pe_0\approx1$}. At even higher pulsation frequencies (\mbox{$De\geq0.7$}), all experiments exhibit a measurable residual birefringence, regardless of how small the strain rate amplitude is within the explored \mbox{$Pe_0-De$} space.

\subsubsection{Lissajous curves}\label{sec_Lissa}
\begin{figure*}[ht]
\centering
\includegraphics[width=\textwidth]{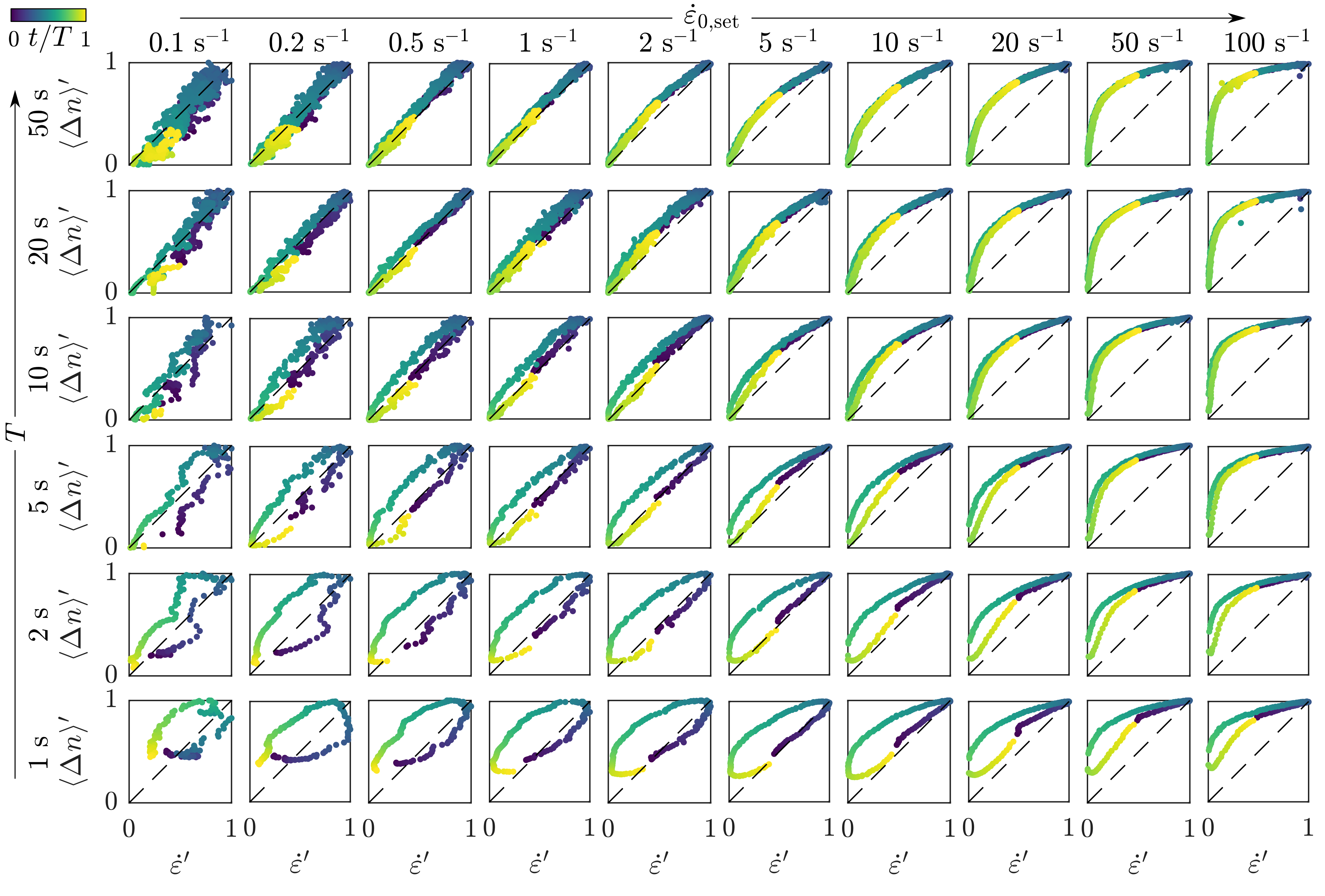}
\caption{Experimental Lissajous curves of the normalized average birefringence \mbox{$\langle\Delta n \rangle^\prime=\langle\Delta n \rangle(t)/\langle\Delta n \rangle_{\mathrm{max}}$} as a function of the normalized extension rate \mbox{$\dot{\varepsilon}^\prime=  \dot{\varepsilon}(t)/\dot{\varepsilon}_{\mathrm{max}}$} for various \mbox{$T$} and \mbox{$\dot{\varepsilon}_{\mathrm{{0,set}}}$}. Data is shown for the \mbox{$\unit[0.1]{\%}$} sample. Black dashed lines correspond to slopes of 1.}
\label{FIG_LissaEXP}
\end{figure*}

To analyze the temporal evolution of flow-induced birefringence during the cycle, we examine the system’s response to sinusoidal strain rate modulation. Figure~\ref{FIG_LissaEXP} presents Lissajous figures of the normalized average birefringence \mbox{$\langle\Delta n \rangle^\prime = \langle\Delta n \rangle(t)/\langle\Delta n \rangle_{\mathrm{max}}$} plotted against the normalized extension rate \mbox{$\dot{\varepsilon}^\prime = \dot{\varepsilon}(t)/\dot{\varepsilon}_{\mathrm{max}}$}, for various values of \mbox{$T$} and \mbox{$\dot{\varepsilon}_{\mathrm{0,set}}$}, representatively shown for the \mbox{$\unit[0.1]{\%}$} CNC sample.

At large LAOE periods and low strain rate amplitudes, \textit{e.g.}, \mbox{$T=\unit[50]{s}$} and \mbox{$\dot{\varepsilon}_{\mathrm{0,set}}\leq\unit[1]{s^{-1}}$} in Fig.~\ref{FIG_LissaEXP}, the birefringence increases linearly with the strain rate, as indicated by the black dashed lines. Keeping \mbox{$T=\unit[50]{s}$} fixed and increasing the strain rate amplitude results in a bent Lissajous curve. As the instantaneous strain rate increases during the cycle, the flow-induced birefringence begins to saturate, resembling the behavior under steady flow conditions (see Fig.~\ref{FIG_SteadyFIB}(b–c)). This indicates a transition from a linear to a nonlinear (curved) Lissajous figure.

When the strain rate amplitude is kept low, \textit{e.g.}, \mbox{$\dot{\varepsilon}_{\mathrm{0,set}}=\unit[0.1]{s^{-1}}$} in Fig.~\ref{FIG_LissaEXP}, and the pulsation frequency is increased, the Lissajous curves evolve from straight lines into pronounced loops. Simultaneously, the birefringence no longer decays to zero but instead exhibits a residual component, as discussed in relation to  Fig.~\ref{FIG_Residual}.

At higher strain rate amplitudes, \textit{e.g.}, \mbox{$\dot{\varepsilon}_{\mathrm{0,set}}\geq\unit[10]{s^{-1}}$} in Fig.~\ref{FIG_LissaEXP}, the Lissajous curves become increasingly open and form pronounced hysteresis loops. The magnitude of this hysteresis grows with increasing pulsation frequency.

Figure~\ref{FIG_LissaEXP} thus demonstrates that hysteresis loops in the Lissajous plots of \mbox{$\langle\Delta n \rangle^\prime$} versus \mbox{$\dot{\varepsilon}^\prime$} emerge as the LAOE period decreases. This indicates that the birefringence response during the ascending and descending phases of the sinusoidal cycle becomes asymmetric at sufficiently high frequencies and amplitudes (see Figs.~\ref{FIG_PulseEXP} and \ref{FIG_PulseSIM}). 

The Lissajous curves of \mbox{$\langle\Delta n \rangle^\prime$} versus \mbox{$\dot{\varepsilon}^\prime$} exhibit distinctive nonlinear loop shapes, reminiscent of those reported in large-amplitude oscillatory uniaxial extension of various complex fluids. In the work of Dessi~\textit{et al.},\cite{Dessi2017} elastomers displayed nonlinear stress–strain behavior attributed to extension-induced thickening, characterized by convex or concave banana-like patterns in Lissajous plots of stress versus strain. Comparable looped nonlinear Lissajous figures have also been observed in soft polymer networks and polymer melts.\cite{Rasmussen2008, Bejenariu2010} More recent studies investigated dilute polymer solutions under LAOE, revealing a nonlinear dependence of stress on strain rate at elevated deformation amplitudes.\cite{Recktenwald2025}

In contrast to these cases, where nonlinearity arises from viscoelastic effects, the curved Lissajous figures observed in this work (see Fig.~\ref{FIG_LissaEXP}) are attributed to saturation of the birefringence at high instantaneous strain rates. Since the CNC suspensions studied here do not exhibit pronounced viscoelasticity under extension, the observed nonlinear behavior reflects optical saturation rather than viscoelastic stress response.

\begin{figure}
\centering
\includegraphics[width=0.45\textwidth]{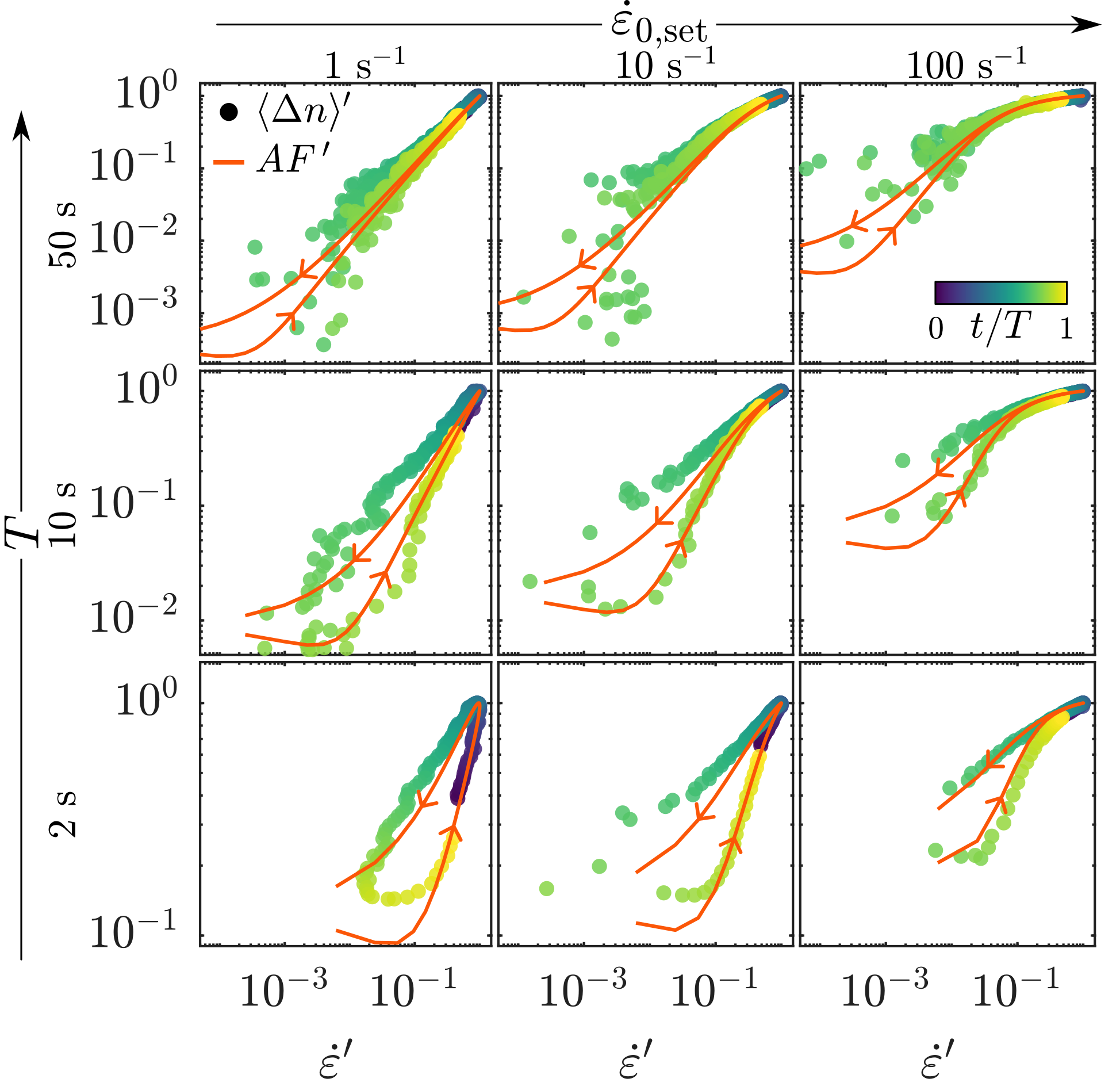}
\caption{Lissajous curves of the normalized average birefringence \mbox{$\langle\Delta n \rangle^\prime$} from the experiments and the normalized anisotropy factor \mbox{$AF^\prime = AF(t)/AF_{\mathrm{max}}$} from the polydisperse simulation based on $A_3$ are plotted as a function of the normalized extension rate \mbox{$\dot{\varepsilon}^\prime$}. Experimental data are shown for the \mbox{$\unit[0.1]{\%}$} sample. Data is represented in log-log scale at representative values of \mbox{$T$} and \mbox{$\dot{\varepsilon}_{\mathrm{0,set}}$}.}
\label{FIG_LissaSIM} 
\end{figure}

Furthermore, we compare the experimental Lissajous curves with numerical results from the polydisperse simulations (\mbox{$A_3$}), which showed the best agreement under both steady and time-dependent flow conditions (see Figs.\ref{FIG_SteadySIM} and \ref{FIG_PulseSIM}). Figure~\ref{FIG_LissaSIM} presents representative Lissajous curves for the \mbox{$\unit[0.1]{\%}$} sample, overlaid with the corresponding normalized anisotropy factor \mbox{$AF^\prime = AF(t)/AF_{\mathrm{max}}$} from the simulations. The data are shown in log-log scale to highlight the behavior as \mbox{$\dot{\varepsilon} \rightarrow 0$}.

Overall, we observe good qualitative agreement between simulations and experiments. The simulations clearly exhibit loops in the Lissajous curves of the anisotropy factor versus strain rate. While the experimental measurements are constrained by the polarization camera’s lower detection limit at low birefringence values, the numerical simulations suggest that hysteresis also occurs at low LAOE frequencies and strain rate amplitudes, \textit{e.g.}, \mbox{$T=\unit[50]{s}$} and \mbox{$\dot{\varepsilon}_{\mathrm{0,set}}=\unit[1]{s^{-1}}$} in Fig.~\ref{FIG_LissaSIM}. However, the corresponding birefringence values lie well below the experimentally accessible range.

\section{Conclusions}
This study investigates the dynamics of model colloidal rod-like systems under large amplitude oscillatory extension (LAOE). A microfluidic OSCER device was used to generate a homogeneous planar extensional flow, through which the extension rate was modulated sinusoidally using precision programmable syringe pumps. Our experimental rod-like colloidal system is based on dilute suspensions of well-characterized cellulose nanocrystals (CNC). The time-dependent strain rate within the OSCER was characterized using micro-particle image velocimetry. The orientation of the colloidal rods in response to the time-dependent strain rate was determined using flow-induced birefringence imaging, performed with a high-speed polarization camera.

Under steady flow conditions, the birefringence increases with strain rate until saturating above \mbox{$Pe \gtrsim 10$} as the CNC rods become highly aligned. This behavior is consistent with previous studies on steady planar extension of CNC suspensions and is also well captured by both our monodisperse and polydisperse numerical simulations, with the rotational diffusion coefficient matched to the experimental system.

In LAOE, the birefringence deviates from a sinusoidal waveform at higher strain rate (or P\'{e}clet) amplitude due to intra-cycle saturation of the particle alignment at high instantaneous strain rates. This behavior is also reproduced in the simulations, with the polydisperse case \mbox{$A_3$} (which is the most strongly weighted to the longest rods in the CNC distribution) showing the best overall qualitative agreement with the experimental data. We demonstrate that hysteresis loops in the Lissajous plots of \mbox{$\langle\Delta n \rangle^\prime$} versus \mbox{$\dot{\varepsilon}^\prime$} emerge as the LAOE period decreases (\textit{i.e.}, as the Deborah number increases). In addition, a residual birefringence that does not decay to zero during the cycle arises at sufficiently high Deborah numbers. These high $De$ effects are due to the rods having insufficient time to equilibrate in the rapidly varying strain rate.  

We believe that these original results should provide a firm foundation for understanding, predicting, and optimizing colloidal particle alignment in processing applications involving complex transient extensional and mixed flows.

\textbf{Acknowledgements:} SMR, VC, AQS, and SJH gratefully acknowledge the support of Okinawa Institute of Science and Technology Graduate University (OIST) with subsidy funding from the Cabinet Office, Government of Japan, along with additional financial support from the Japanese Society for the Promotion of Science (JSPS, Grant Nos. 24K07332, 24K17736, and 24K00810). VC also acknowledges support from the ``la Caixa'' Foundation (ID 100010434), fellowship code LCF/BQ/PI25/12100026. The authors thank Dr. Yuto Yokoyama from OIST for fruitful discussions.
\vspace{-0.02in}

\footnotesize{
%\bibliography{reference} %your .bib file
\providecommand*{\mcitethebibliography}{\thebibliography}
\csname @ifundefined\endcsname{endmcitethebibliography}
{\let\endmcitethebibliography\endthebibliography}{}

}

\end{document}